\documentclass[prb,onecolumn,nofootinbib,citeautoscript,eqsecnum,10pt,notitlepage]{revtex4-2}
\pdfoutput=1

\usepackage{float}
\usepackage{array}
\usepackage{slashed,bbold}

\usepackage{amsmath,amssymb,bm} 
\usepackage{graphicx}

\usepackage[tight]{subfigure} 

\usepackage{color} 
\usepackage[papersize={8.5in,11in}]{geometry}

\usepackage{color}
\definecolor{darkblue}{rgb}{0.,0.,0.4}
\definecolor{darkred}{rgb}{0.5,0.,0.}
\definecolor{BlueViolet}{RGB}{138,43,226}
\definecolor{SkyBlue}{RGB}{30,144,255}
\definecolor{DarkGreen}{RGB}{0,100,0}
\usepackage[pdftex,colorlinks=true,linkcolor=darkblue,citecolor=blue,urlcolor=darkred]{hyperref}

\def \nn{\nonumber \\}

\begin{document}

\title{Transmission in pseudospin-1 and pseudospin-3/2 semimetals with linear dispersion through scalar and vector potential barriers}

\author{Ipsita Mandal}

\affiliation{Faculty of Science and Technology, University of Stavanger, 4036 Stavanger, Norway}
\affiliation{Nordita, Roslagstullsbacken 23, SE-106 91 Stockholm, Sweden}

\begin{abstract}
We investigate the tunneling of pseudospin-1 and pseudospin-3/2 quasiparticles through a barrier consisting of both electrostatic and vector potentials, existing uniformly in a finite region along the transmission axis. First, we find the tunneling coefficients, conductivities, and Fano factors in the absence of the vector potential. Then we repeat the calculations by switching on the relevant magnetic fields. The features show clear distinctions, which can be used to identify the type of semimetals, although both of them exhibit linear band-crossing points.
\end{abstract}
\maketitle

\tableofcontents

\section{Introduction}
 Recently, there has been a surge of interest in condensed matter systems that can host multiband (both linear and quadratic) crossings in the Brillouin zone (BZ) \cite{bernavig}, many of which do not have a high-energy counterpart. In particular, for threefold as well as for a class of  fourfold  degeneracies, the low energy Hamiltonian is of the form $\mathbf{k} \cdot \mathbf{\mathcal{S}}$,
 where $\mathbf{\mathcal{S}}$ represents the vector consisting of three spin-1 or spin-3/2 matrices.
Hence, we get three-dimensional (3d) semimetals with pseudospin-1 and pseudospin-3/2 quasiparticle excitations, which are nothing but natural generalizations of the Weyl semimetal Hamiltonian $\mathbf{k} \cdot \mathbf{\sigma}$ ($\mathbf{\sigma}$ representing the vector of the Pauli matrices) featuring pseudospin-1/2 quasiparticles. All these fermions have a linear dispersion, just like Dirac fermions, and the bandstructures have nonzero Chern numbers. The pseudospin-1 quasiparticles are sometimes referred to as Maxwell fermions \cite{PhysRevA.96.033634}, while the pseudospin-3/2 quasiparticles are well-known as Rarita-Schwinger-Weyl fermions \cite{long}. By using DFT calculations and bulk-sensitive soft x-ray ARPES, B.~Q.~Lv {\textit{et al}} \cite{lv} have predicted the coexistence of all these three types of topological fermions in the electronic structure  of  PdBiSe.
There is a crucial difference in the dynamical properties of the Dirac particles (with spin-1/2) and the spin-1 quasiparticles that we consider here -- the latter exhibit super-Klein tunneling \cite{urban,lai,zhu}, which means that the barrier is completely transparent for all incident angles for certain incident energies. Note that both Dirac and spin-3/2 particles \cite{peng-he} exhibit Klein tunneling.

In this paper, we study the behaviour of the transmission coefficients of the pseudospin-1 and pseudospin-3/2 fermions in presence of finite barriers made of scalar and vector potentials. We try to identify the distinct features peculiar to the pseudospin value. These might prove to be a tool to identify/distinguish these materials in experiments. Tunneling in 2d optical lattice versions of pseudospin-1 and pseudospin-3/2 fermions have been studied earlier in Ref.~\cite{shen,lan}.

The paper is organized as follows. In Sec.~\ref{formalism}, we explain the general set-up for carrying out the tunneling experiment. In Sec.~\ref{secspin1} and ~\ref{secspin32}, we apply the Landau-Büttiker formalism to compute the tunneling coefficients for the pseudospin-1 and pseudospin-3/2 fermions, respectively.
Finally, we end with a summary and outlook in Sec.~\ref{secsum}.

\section{Formalism}
\label{formalism}

\begin{figure}[htb]
{\includegraphics[width = 0.7 \textwidth]{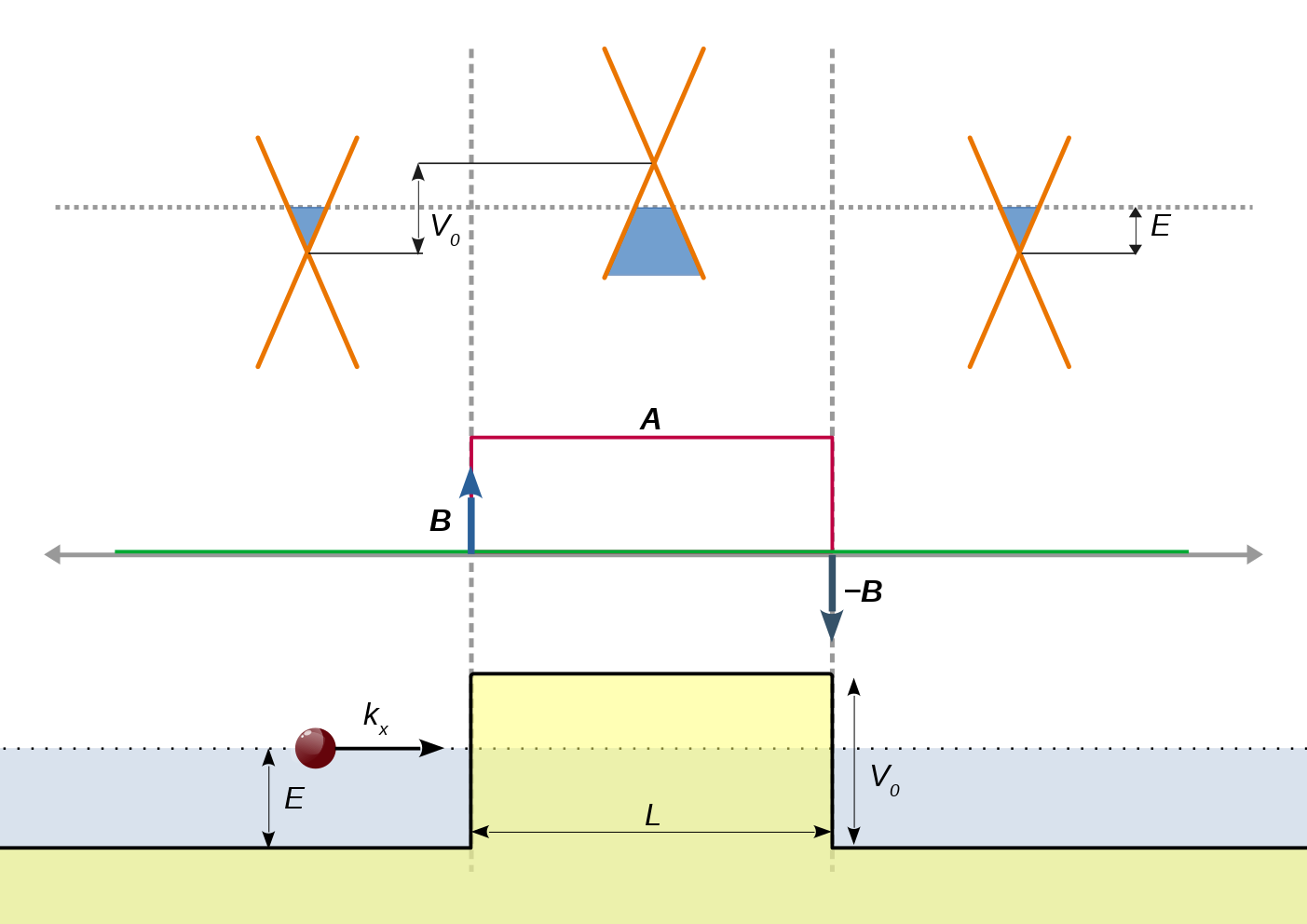}}
\caption{Tunneling through a potential barrier in a linear band-crossing semimetal. The upper panel shows the schematic diagrams of the spectrum of a pair of particle-hole symmetric bands, with respect to a scalar (or electric) potential barrier of strength $V_0$ in the $x$-direction. The middle panel shows a constant vector potential $\mathbf{A}$ superposed
in the same region. Theoretically, this vector potential can be created by applying equal and opposite delta function magnetic fields ($\mathbf{B}$ and $-\mathbf{B}$) at the edges of the barrier region, oriented perpendicular to the $x$-axis.
The lower panel represents the schematic diagram of the transport across the potential barrier. The Fermi level is depicted by dotted lines, and lies in the conduction band outside the barrier, and in the valence band inside it. The blue fillings indicate occupied states. For simplicity, only one pair of particle-hole symmetric bands has been shown. Generically, there can be more than one such pair.
}
\label{figbands}
\end{figure}

In order to study transport, the 3d system is modulated by a scalar potential barrier (giving rise to an electric field) of
strength $V_0 $ and width $L$, resulting in an $x$-dependent potential energy function:
\begin{align}
V ( x ) = \begin{cases} V_0 &  \text{ for } 0 < x < L \\
0 & \text{ otherwise} \,.
\end{cases}
\label{eqpot}
\end{align}

In the next step, we subject the sample to equal and opposite magnetic fields localized at the edges of the rectangular electric potential, and directed perpendicular to the $x$-axis \cite{mansoor,kai}. This can be theoretically modeled as Dirac delta functions of opposite signs at $x=0$ and $x=L$ respectively, and gives rise to a vector potential with the components:
\begin{align}
\mathbf {A}( x ) \equiv  \lbrace 0, A_y, A_z  \rbrace = \begin{cases} 
\lbrace 0, B_z, -B_y  \rbrace &  \text{ for } 0 < x < L \\
\mathbf 0 & \text{ otherwise} \,.
\end{cases}
\label{eqvecpot}
\end{align}
Note that this arises from the magnetic field $\mathbf{B}  = 
\frac{1}{2}
\left(  B_y\, \hat{\mathbf j}  + B_z\, \hat{\mathbf k}   \right )
\left [  \delta\left(x= 0\right) - \delta\left( x=L \right) \right ] $.
The entire set-up is depicted pictorially in Fig.~\ref{figbands}.
Some possible methods to achieve this set-up in real experiments (for instance, by placing ferromagnetic stripes at barrier boundaries) have been discussed
in Ref.~\cite{mansoor}.

We will follow the usual Landau-B\"{u}ttiker procedure (see, for example Refs.~\cite{salehi,beenakker,ips-qbcp-tunnel}) to compute the transport coefficients. For the sake of completeness, we review the important steps here.
We consider the tunneling of quasiparticles in a slab of square cross-section (without any loss of generality), with the transverse width being $W$. We assume that $W$ is large enough such that the specific boundary conditions being used in the calculations are irrelevant for the bulk response. Here, we impose the periodic boundary conditions:
\begin{align}
  \Psi^{\mathrm{tot}}(x,0,z) =
\Psi^{\mathrm{tot}}(x,W,z)\,,\quad
 \Psi^{\mathrm{tot}}(x,y,0) = \Psi^{\mathrm{tot}}(x,y,W) \,.
\end{align}
The  transverse momentum $ \mathbf k_\perp=(k_y,k_z)$ is conserved as no potential is applied along those directions, and its components are quantized as:
\begin{align}
k_y=\frac{ 2\,\pi\,n_y} {W} \equiv q_{n_y} \,,\quad k_z =\frac {2\,\pi\,n_z} {W} \equiv q_{n_z}\,,
\end{align}
where $(n_x, n_y) \in \mathbb{Z}$.
The longitudinal direction corresponds to transport along the $x$-axis, and for this we need to consider plane wave solutions of the form $ e^{\mathrm{i}\,k_x x} $. Then the full wavefunction is given by:
\begin{align}
 \Psi^{\mathrm{tot}} (x,y,z,\mathbf n) =
\text{const.}
\times   \Psi_{ \mathbf{n}}(x)\,  e^{ \mathrm{i}\left( q_{n_y} y +  q_{n_z} z\right ) }\,,
\end{align}
with
\begin{align}
\mathbf{n} =(n_y, n_z)\,.
\end{align}

Since we consider transmission in semimetals with at least one pair of valence ($\varepsilon^-$) and conduction ($\varepsilon^+$) bands crossing linearly at a point, with dispersion relation of the form $\varepsilon^\pm = \pm \hbar \, v_g \sqrt{k_x^2 +k_y^2+k_z^2}$ ($v_g$ is the effective speed of the quasiparticles),
we will deal with the case when the incident particles are electron-like excitations.
In other words, the Fermi energy ($E$) is adjusted to lie in the conduction band outside the potential barrier \footnote{ The Fermi energy $E$ can in general be tuned by chemical doping
or a gate voltage.}.
Hence, given an arbitrary mode of transverse momentum $\mathbf{k}_\perp$, we can determine the $x$-component of the wavevectors of the incoming, reflected, and transmitted waves (denoted by $k_{\ell}$), by solving $ \varepsilon^+(k_x ,  \mathbf n) = E\,.$
In the regions $x<0$ and $x>L$, we have only propagating modes ($  k_\ell $ is real), while the $x$-components
in the scattering region (denoted by $ \tilde k  $),  are given by
$\tilde k^2 = \left( \frac{ E-V_0} {\hbar\,v_g} \right)^2- 
\left(  \mathbf k_{\perp} + \frac{e\,\mathbf A}{\hbar} \right)^2 $,
and may be propagating ($\tilde k $ is real) or evanescent ($\tilde k $ is imaginary).

Now we need to use the piece-wise solutions for the wavefunction ($\Psi$), applicable in the regions in question (inside or outside the potential barrier). Hence, even though the incident wavefunction represents an electron-like excitation, for $V_0> E$, the Fermi level within the potential barrier lies within the valence band, and we must use the valence band wavefunctions (representing hole-like excitations) in that region. In the next step, we need to use the boundary conditions to determine the reflection and transmission coefficients. The boundary conditions are determined by integrating the equation 
$\mathcal{H} \,\Psi = E\, \Psi $ ($\mathcal{H}$ is the Hamiltonian written in the position space) over a small interval in the $x$-direction around the points $x=0$ and $x=L$, and they ensure the continuity of the current flux along the $x$-direction.

\section{Pseudospin-1 Fermions}
\label{secspin1}

It has been shown in Ref.~\cite{bernavig} that the space group $199$ may host a 3d representation at the $P$ point (and its time-reversed partner $-P$) in the BZ, which is time reversal non-invariant. The linearized $\mathbf{k} \cdot \mathbf {p}$ Hamiltonian
about P hosts pseudospin-1 fermions and takes the form:
\begin{align}
\mathcal{H}_1(\mathbf  k) = \hbar\,v_g \,\mathbf{k}\cdot \mathbf S\,,
\end{align}
where $\mathbf S$ represents the vector spin-1 operator with the three components
\begin{align}
S_x = \frac {1} {\sqrt{2}}
\begin{pmatrix}
0&1&0\\1&0&1\\0&1&0
\end{pmatrix}\,,\quad
S_y =\frac{1}{\sqrt{2}}
\begin{pmatrix}
0&-\mathrm{i} &0\\ \mathrm{i} &0&- \mathrm{i}\\
0& \mathrm{i} &0
\end{pmatrix}\,,\quad
S_z =
\begin{pmatrix}
1&0&0\\0&0&0\\0&0&-1
\end{pmatrix}\,,
\end{align}
and $v_g$ denotes the magnitude of the group velocity associated with the Dirac cone. 
The energy eigenvalues are given by:
\begin{align}
\varepsilon_{1}^{\pm}(\mathbf{k}) = \pm  \hbar \, v_g \, k \,,
\quad \varepsilon_{1}^{0}(\mathbf k) = 0\,,
\end{align}
where $k= \sqrt{k_x^2 +k_y^2+k_z^2}$, and demonstrate two linearly dispersing bands and a flat-band crossing at a point.
Here the ``$+$" and ``$-$" signs refer to the linearly dispersing conduction and valence bands, respectively.
The corresponding normalized eigenvectors are given by:
\begin{align}
\Psi_s & = \frac{1} { \mathcal{N}_s }
\left\{\frac{2 k_z \left (k_z +s\, k \right )
+k_x^2+k_y^2}{(k_x+ \mathrm{i}\, k_y)^2},
\frac{\sqrt{2} \left (k_z + s\,k \right )} {k_x+ \mathrm{i}\, k_y},1\right\}^T
 \left( \text{ where } s=\pm \right) \,,\quad
  \Psi_0 = \frac{1} { \mathcal{N}_0 }
 \left\{\frac{-k_x+\mathrm{i}\,  k_y}
  {k_x+ \mathrm{i}\, k_y},\frac{\sqrt{2}\, k_z}{k_x+ \mathrm{i}\, k_y},1\right\},
\end{align}
respectively, where the $\frac{1} { \mathcal{N}_s}$ and $\frac{1} {\mathcal{N}_0 }$ denote the corresponding
normalization factors.

The current operator for this is system is captured by $\hat{\mathbf{j}}= \nabla_k \mathcal{H}_1(\mathbf  k)
= v_g \,\mathbf{S}\,,$ which implies that the local current for a flat-band plane wave is given by:
\begin{align}
\mathbf{j}_0 =v_g  \Psi_0^\dagger \,\mathbf{S} \, \Psi_0 =0\,.
\end{align}
Hence, it does not contribute to the current density \cite{lai}, and we need only consider $\Psi_\pm$ for transport properties.

In presence of the scalar and vector potentials, we need to consider the total Hamiltonian:
\begin{align}
\mathcal{H}_{1}^{tot} &=
 \mathcal{H}_{1}
\left[ -\mathrm{i}\,\nabla + \frac{e\,\mathbf A(x)}{\hbar} \right ]
 +V(x)
\end{align} 
in position space, and find the appropriate wavefunctions.

\subsection{Mode-matching}

A scattering state $\Psi_{ \mathbf n}$, in the mode labeled by $\mathbf n$, is
constructed from the states:
\begin{align}
 \Psi_{ \mathbf n} (x)=&  \begin{cases}   \phi_L & \text{ for } x<0  \\
 \phi_M & \text{ for } 0< x < L \\
  \phi_R &  \text{ for } x > L 
\end{cases} \,,\nonumber \\
  \phi_L = & \,\frac{   
 \Psi_{+} (   k_\ell,  \mathbf{k}_\perp) \, e^{\mathrm{i}\, k_\ell x }
+  
 r_{\mathbf n} \,\Psi_{+} ( -k_{\ell},  \mathbf{k}_\perp) \, e^{-\mathrm{i}\, k_\ell x }
}
{\sqrt{  { \mathcal{V} }(k_\ell, \mathbf n) }}
\, ,\nonumber \\
  \phi_M  = &\,\Big[  
 \alpha_{\mathbf n} \,\Psi_{+} ( \tilde k,  \tilde {\mathbf{k}}_\perp) \,
 e^{\mathrm{i}\,\tilde k\,  x } 
 + 
 \beta_{\mathbf n} \,\Psi_{+} (  -\tilde k, \tilde {\mathbf{k}}_\perp) \,
 e^{-\mathrm{i}\,\tilde k  \, x } 
\Big]   \Theta\left( E-V_0  \right)
 \nonumber\\ & 
 + \Big[
 \alpha_{\mathbf n} \,\Psi_{-} ( \tilde k, \tilde {\mathbf{k}}_\perp) \,
 e^{\mathrm{i}\,\tilde k\,  x } 
 + 
 \beta_{\mathbf n} \,\Psi_{-} (  -\tilde k, \tilde {\mathbf{k}}_\perp) \,
 e^{-\mathrm{i}\,\tilde k  \, x }
   \Big]  \,\Theta\left( V_0-E  \right),\nonumber \\
 \phi_R = & \,\frac{  t_{\mathbf n} \,\Psi_{+} ( k_{\ell},  \mathbf{k}_\perp) 
 } 
{\sqrt{  { \mathcal{V} }  (k_\ell, \mathbf n)}}
\, e^{\mathrm{i}\, k_\ell \left( x-L\right)}\,,\nonumber \\
 { \mathcal{V} }(k_\ell, \mathbf n) \equiv &  \, |\partial_{k_\ell} \varepsilon_{1}^+ (k_\ell, \mathbf n)|
= \frac{ \hbar\,v_g  k_\ell } {k}\,,
\quad k_{\ell} = \sqrt{\frac{E^2} {\hbar^2 v_g^2}-  \mathbf{k}_\perp^2}
\,,\quad  
\tilde k = \sqrt{\frac{ \left( E-V_0\right)^ 2} {\hbar^2 v_g^2} 
-{ \tilde {\mathbf{k}}_\perp} ^2 }\,,\quad 
 \tilde {\mathbf{k}}_\perp = \mathbf{k}_\perp +  \frac{e\,\mathbf A(x)}{\hbar} \,,
\end{align}
where we have used the velocity $ { \mathcal{V} }(k_\ell, \mathbf n) $ to normalize the incident, reflected and transmitted plane waves. The symbol $\Theta(u)$ represents the Heaviside step function, as usual.
The mode-matching procedure at the edges $x=0$ and $x=L$ gives us the explicit expressions for
$t_{\mathbf n}(E, V_0,\mathbf B) $, which are too long to write down within the manuscript.
In any case, we have to compute the transmission probability numerically, which at an energy $E$ is given by: 
\begin{align}
T( E ,  V_0,\theta , \phi,\mathbf B) = | t_{\mathbf n}( E, V_0,\mathbf B )|^2 \,,
\text{ where } 
\theta = \cos^{-1} \left( \frac{\hbar\,v_g \,q_{n_z}} { E } \right)
\text{ and }
\phi = \tan^{-1} \left( \frac {q_{n_y}} {k_{\ell,3/2}} \right)
\end{align}
define the incident angle (solid) of the incoming wave in 3d.

At normal incidence, the analytical expression simplifies to $ t_{\mathbf 0}(E, V_0,0)
= e^{\frac{\mathrm{i}\,L\,\left( E-V_0 \right) } {\hbar\,v_g}}$, which results in perfect transmission ($T=1$), also referred to as Klein tunneling. Again, $  t_{\mathbf n}( V_0/2, V_0,\mathbf 0 ) =e^{ \frac{\mathrm{i}\,L \,V_0 \sin \theta \cos\phi}{2\,\hbar\,v_g} }$, which implies the occurrence of perfect transmission for any incident angle when $E=V_0/2$. This is the well-known super-Klein tunneling \cite{lai,zhu} for pseudospin-1 Dirac cone systems. We also note that $  t_{\mathbf n\neq \mathbf 0}( V_0, V_0,\mathbf 0 ) = 0$.

\begin{figure}[htb]
\subfigure[]{\includegraphics[width = 0.14 \textwidth]{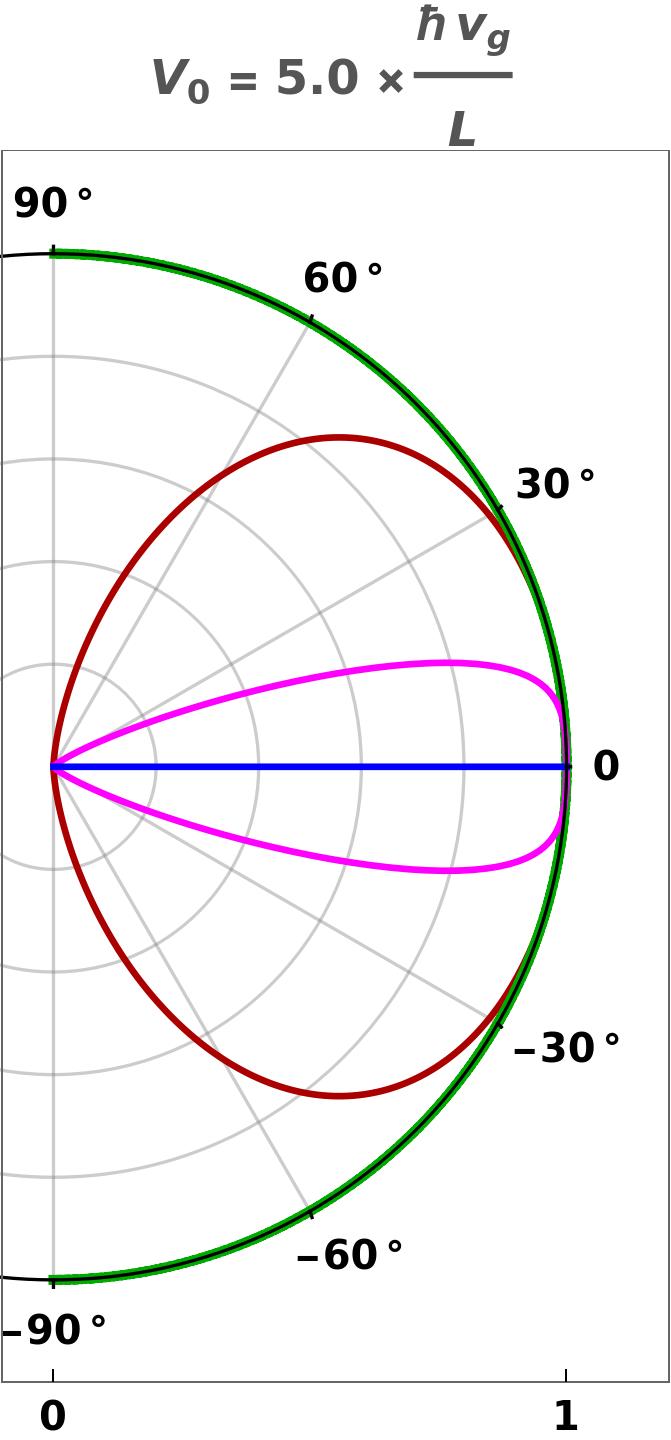}} \quad
\subfigure[]{\includegraphics[width = 0.14 \textwidth]{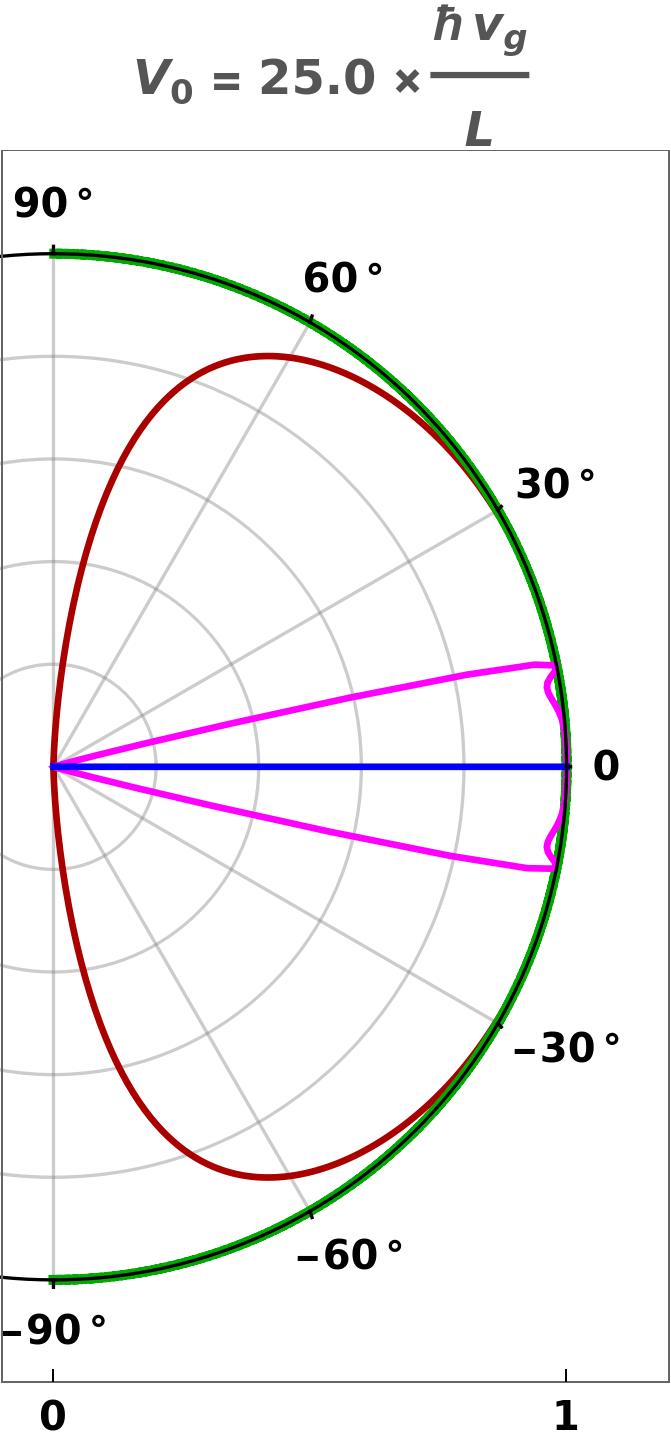}}\quad
\subfigure[]{\includegraphics[width = 0.14 \textwidth]{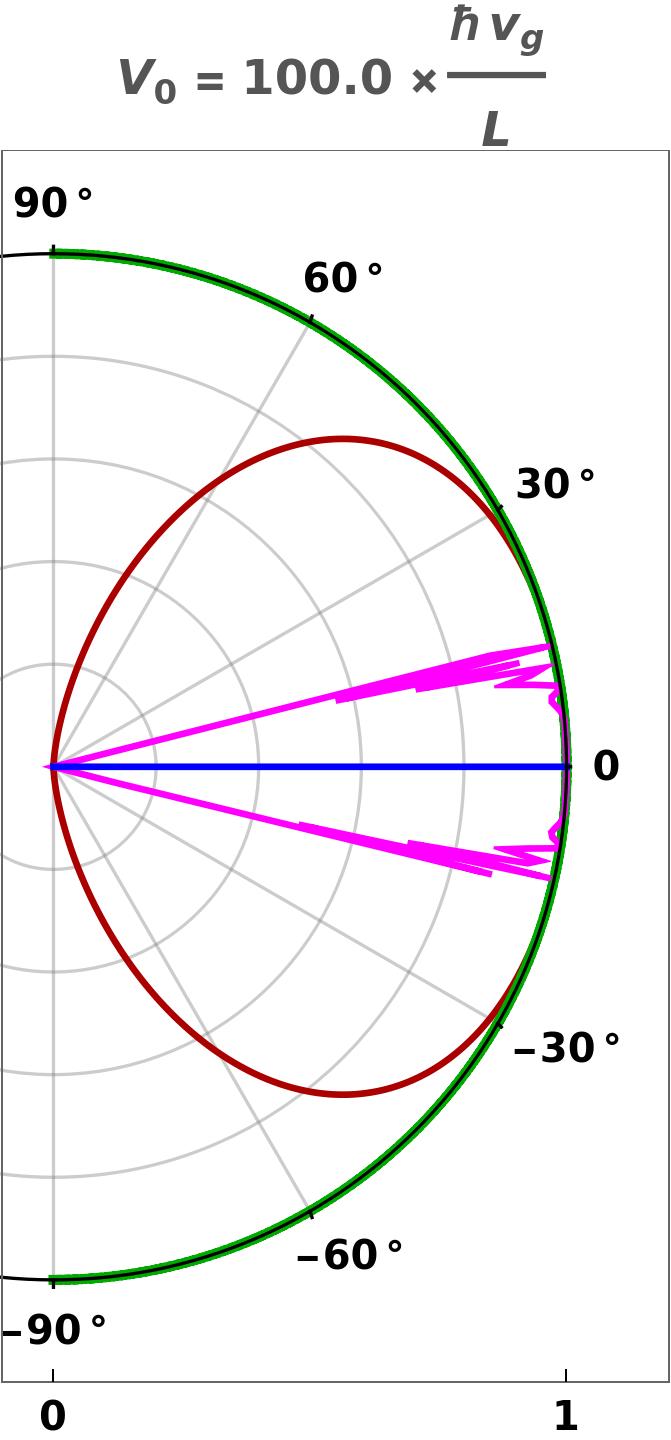}}  \quad
\subfigure[]{\includegraphics[width = 0.14 \textwidth]{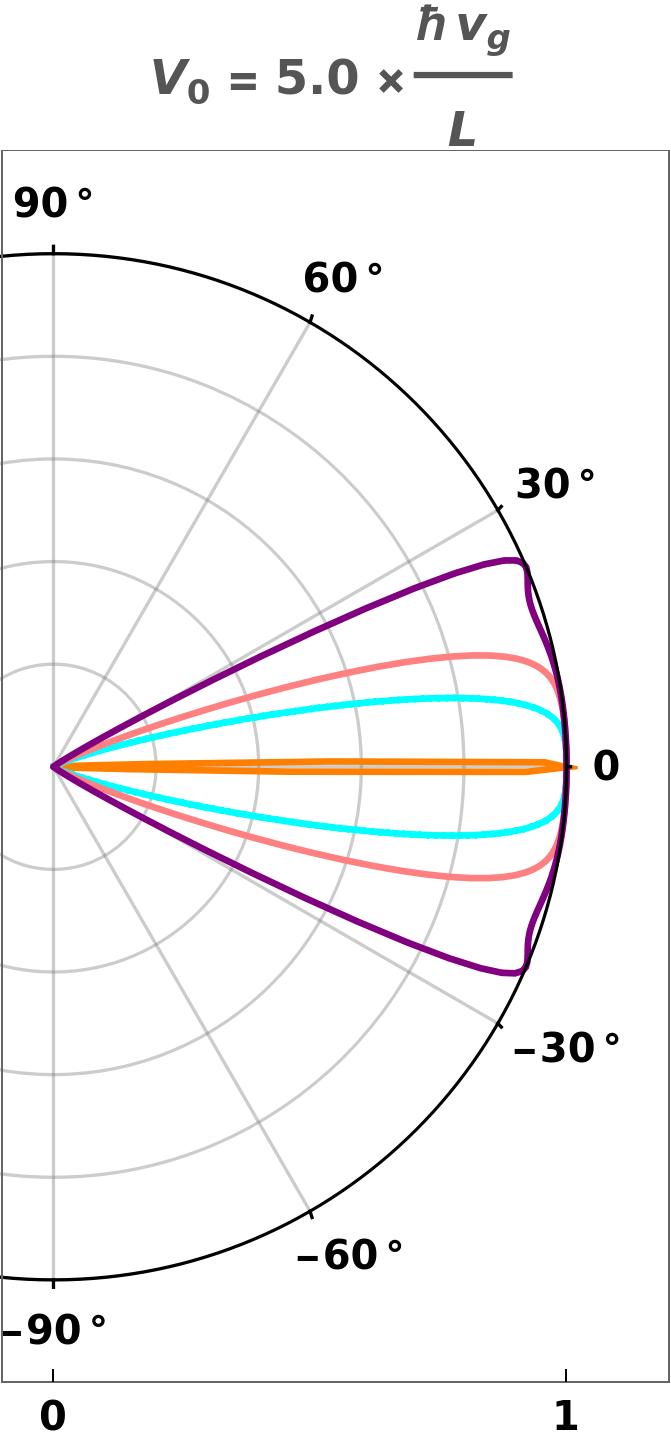}}  \quad
\subfigure[]{\includegraphics[width = 0.14 \textwidth]{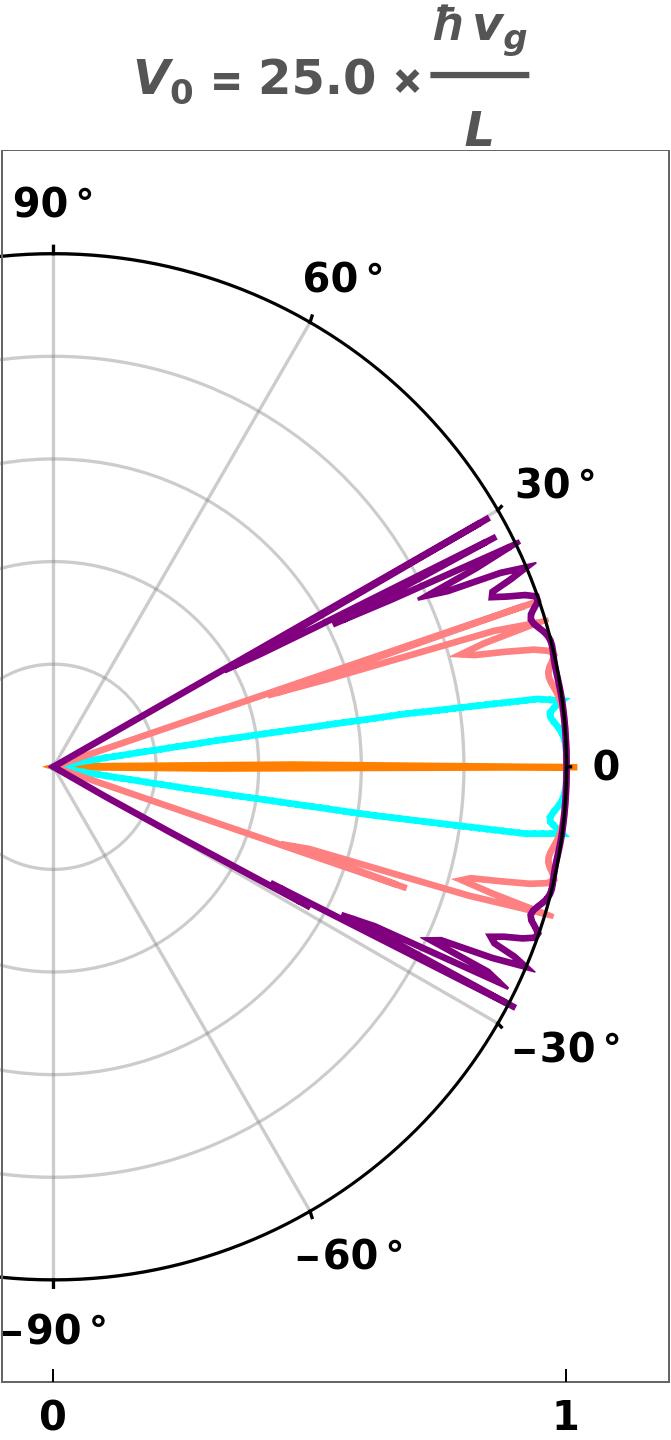}}  \quad
\subfigure[]{\includegraphics[width = 0.14 \textwidth]{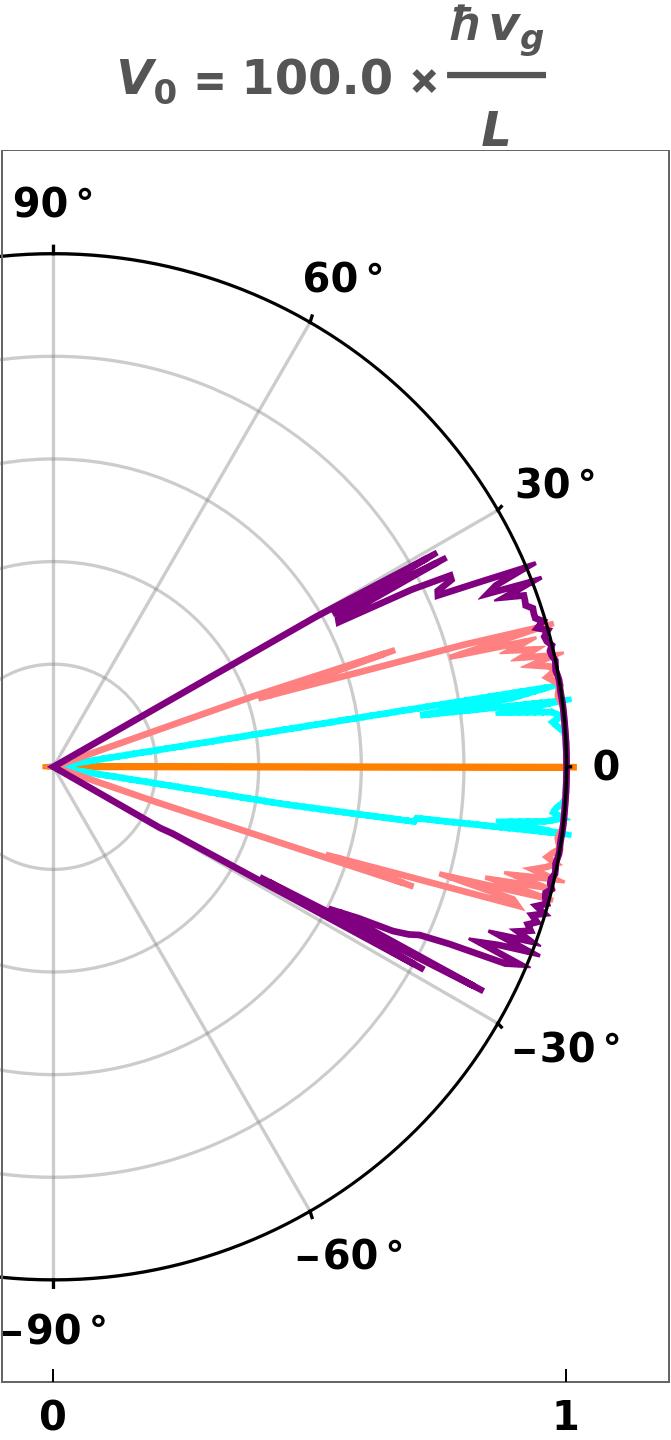}}
\caption{Pseudospin-1 semimetal: The polar plots show the transmission coefficient $T(E, V_0,\theta,\frac{\pi}{2},\mathbf 0) $ as a function of the incident angle $\phi$ (in the $xy-$plane with no $k_z-$component) for the parameters $E= 0.3 \,V_0$ (red), $E=0.5\,V_0$ (green), $E=0.8\,V_0$ (magenta), $E= V_0$ (blue), $E=1.001\,V_0$ (orange), $E=1.2 \,V_0$ (cyan), $E=1.5\,V_0$ (pink), and $E=2.0 \,V_0$ (purple). Super-Klein tunneling manifests itself at $E=V_0/2$, for which $T=1$.}
\label{figpolarspin1}
\end{figure}

\subsection{Transmission coefficient, conductivity, and Fano factor}

\begin{figure}[htb]
\subfigure[]
{\includegraphics[width = 0.23 \textwidth]{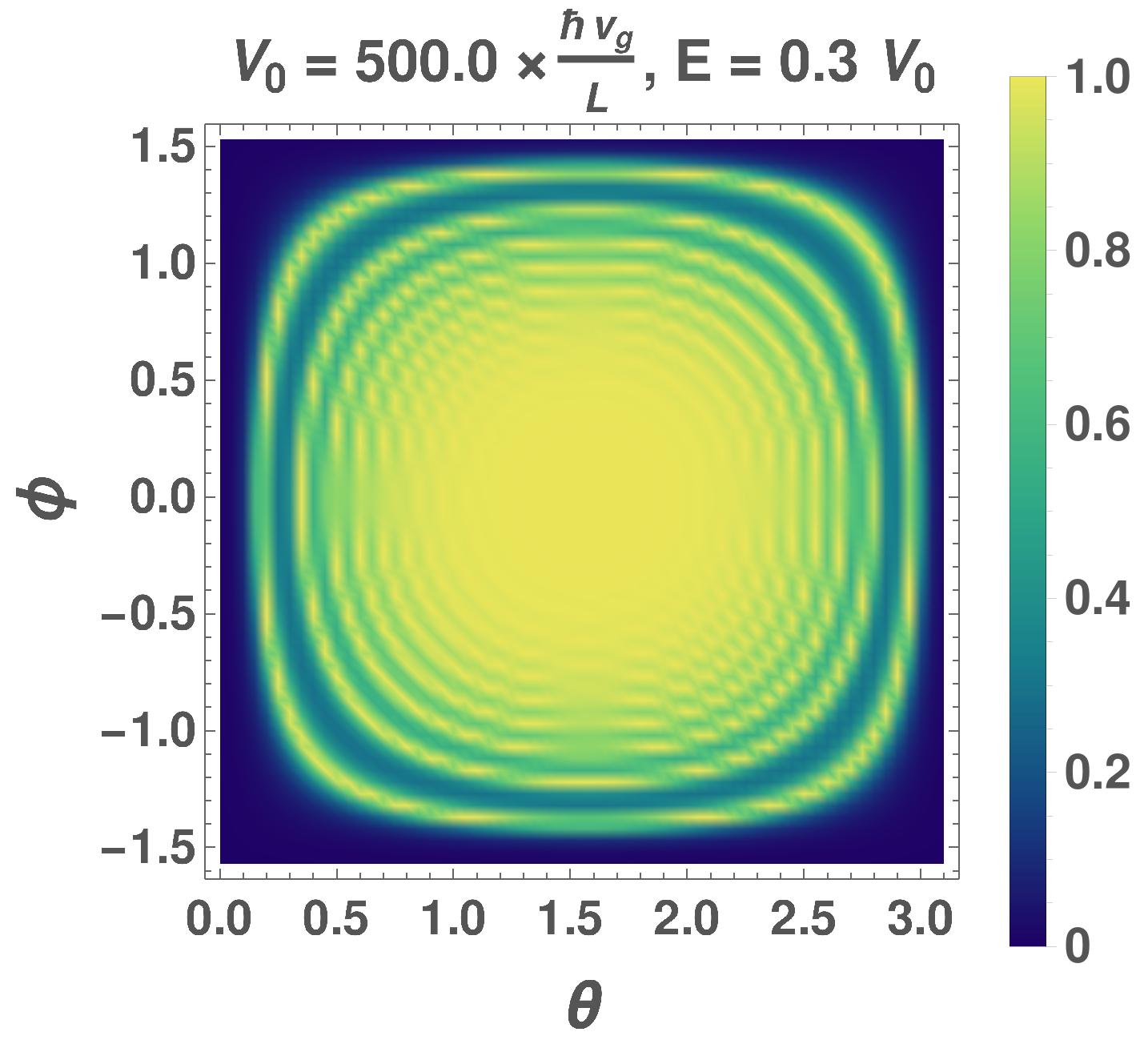}} \quad
\subfigure[]
{\includegraphics[width = 0.23 \textwidth]{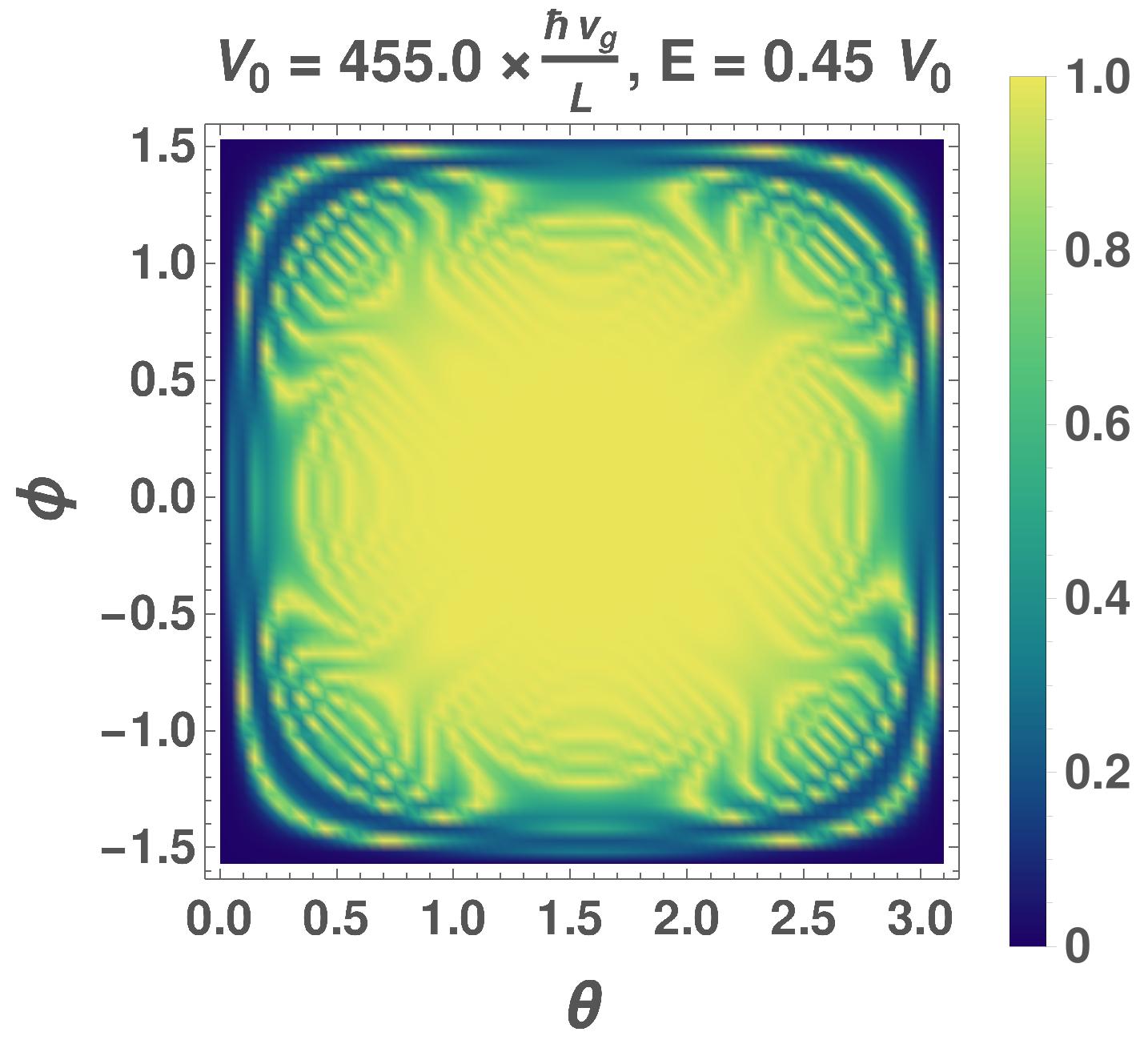}} \quad
\subfigure[]
{\includegraphics[width = 0.23 \textwidth]{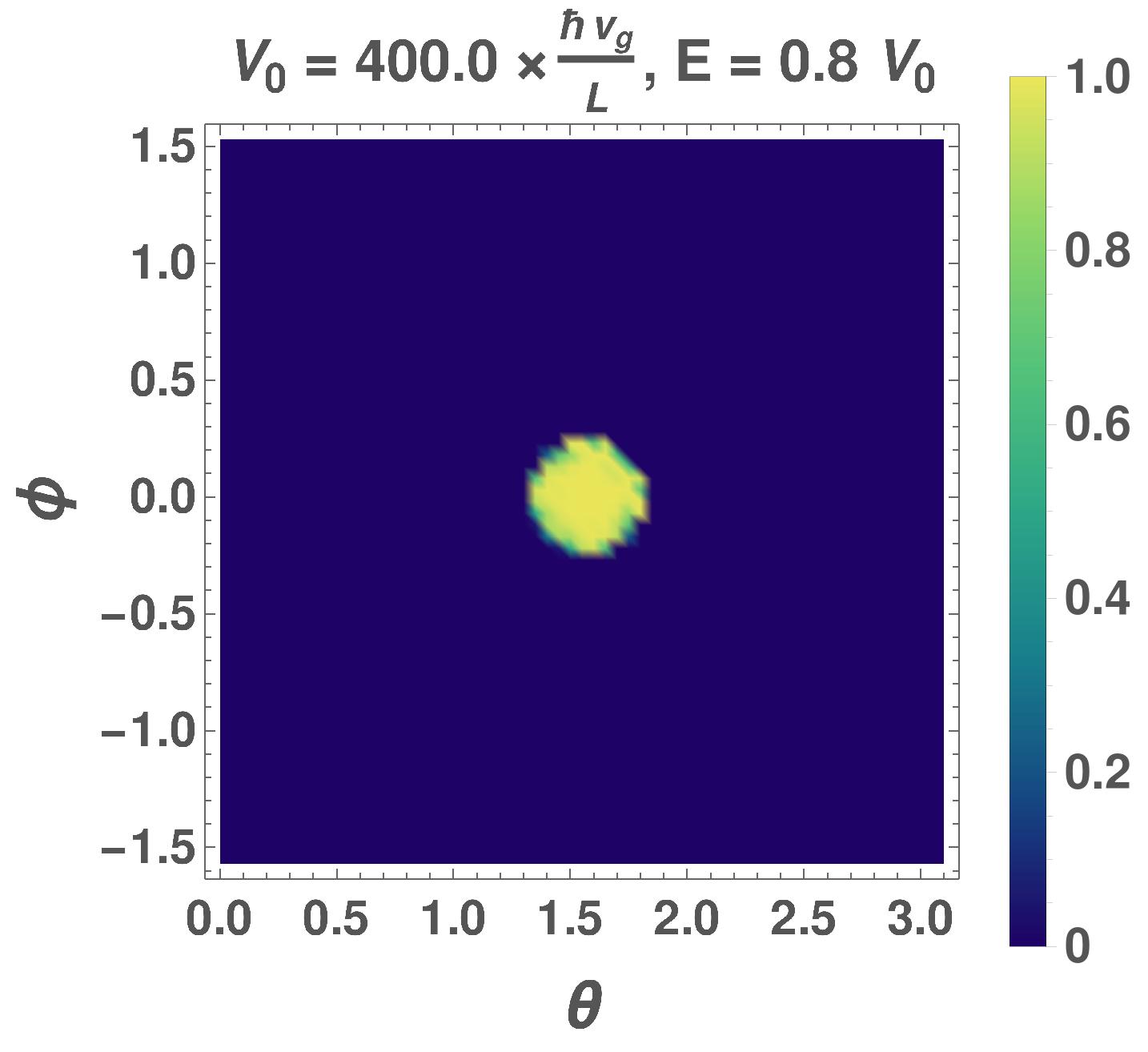}}\quad
\subfigure[]
{\includegraphics[width = 0.23 \textwidth]{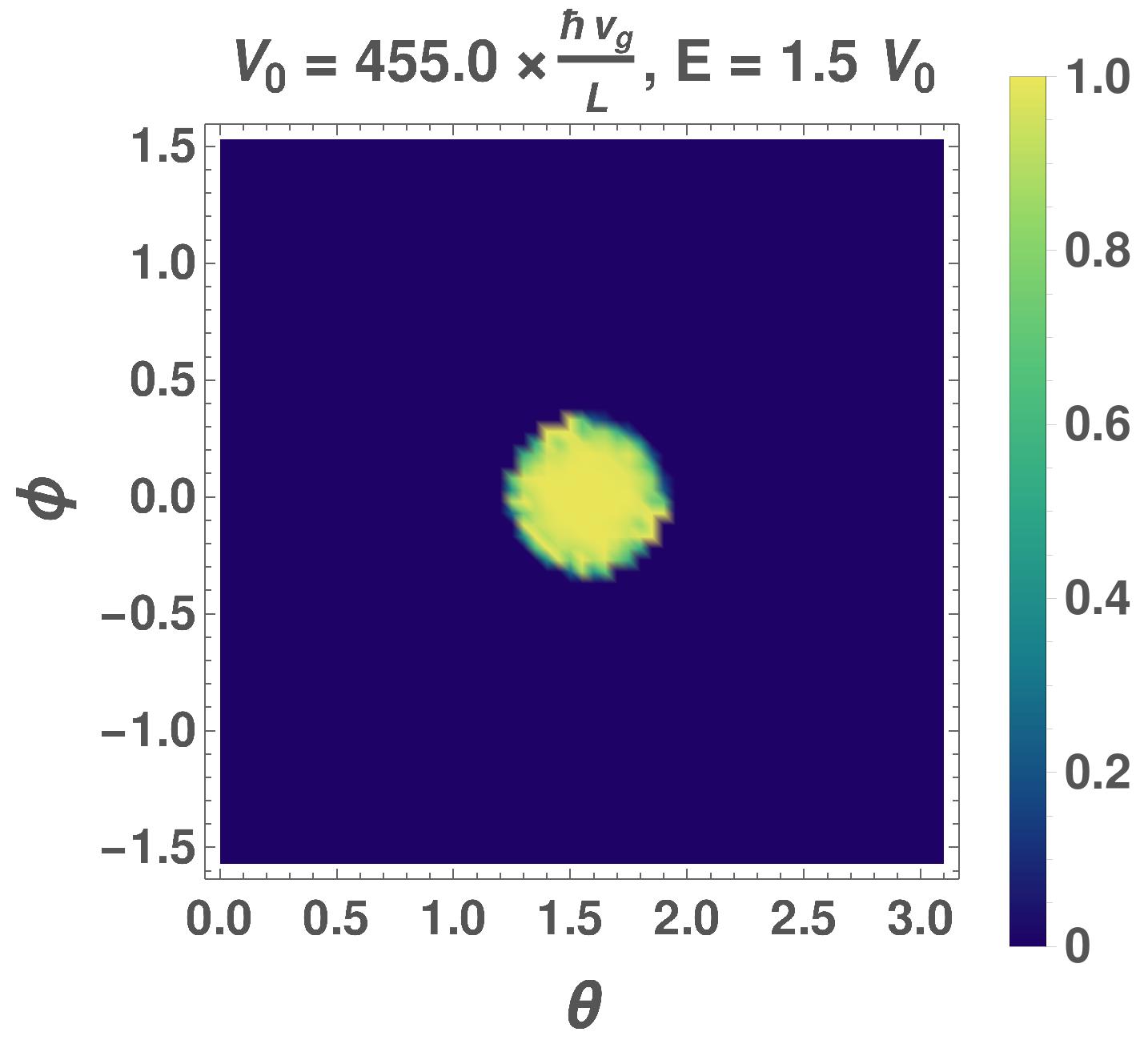}} 
\caption{Pseudospin-1 semimetal: Contourplots of the transmission coefficient ($T$) in the absence of the vector potential, as a function of $(\theta, \phi)$, for various values of $V_0$ and $E$.}
\label{figcontourspin1}
\end{figure}

\begin{figure}[htb]
\subfigure[]
{\includegraphics[width = 0.45 \textwidth]{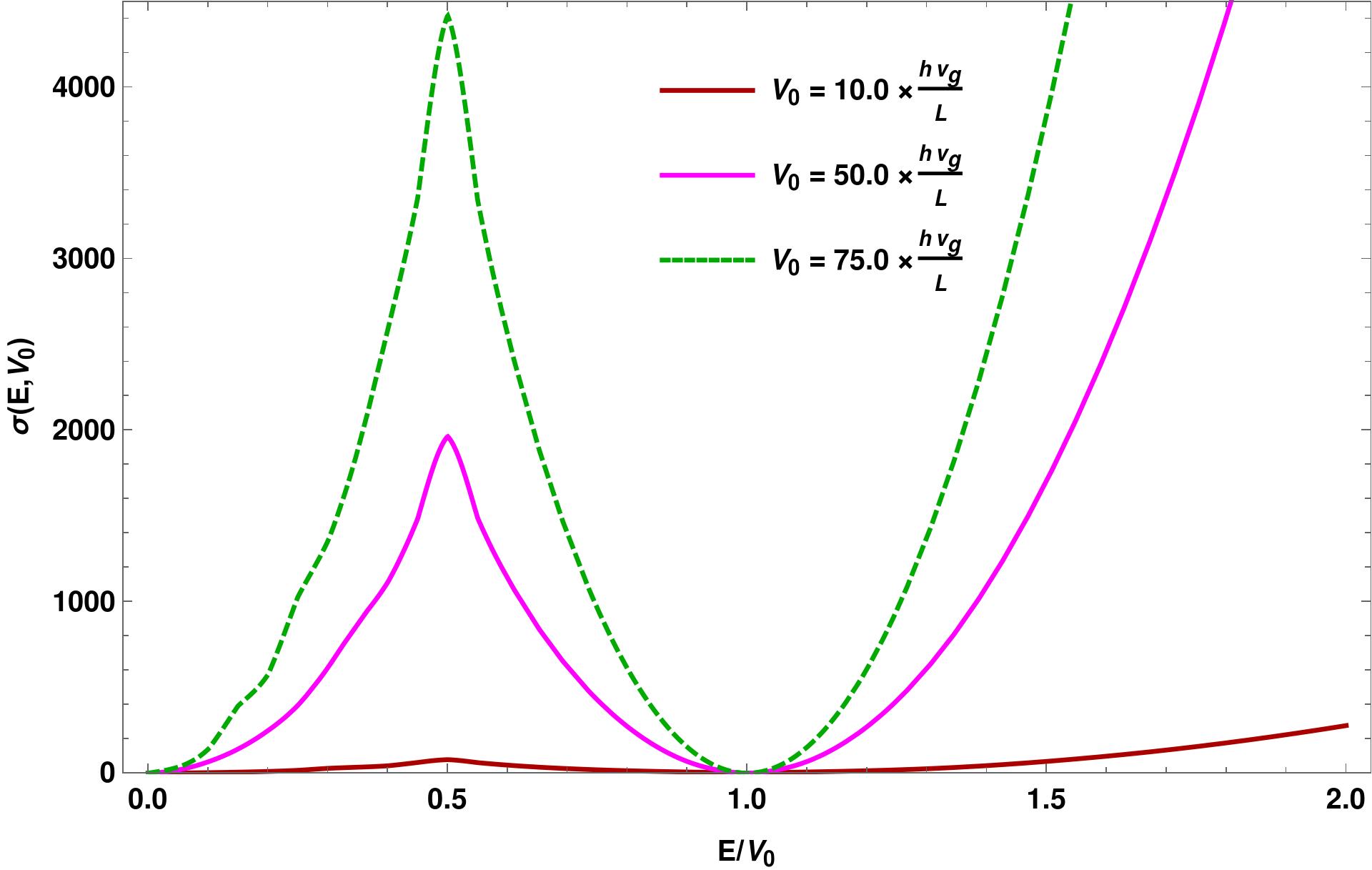}}\qquad
\subfigure[]
{\includegraphics[width = 0.45 \textwidth]{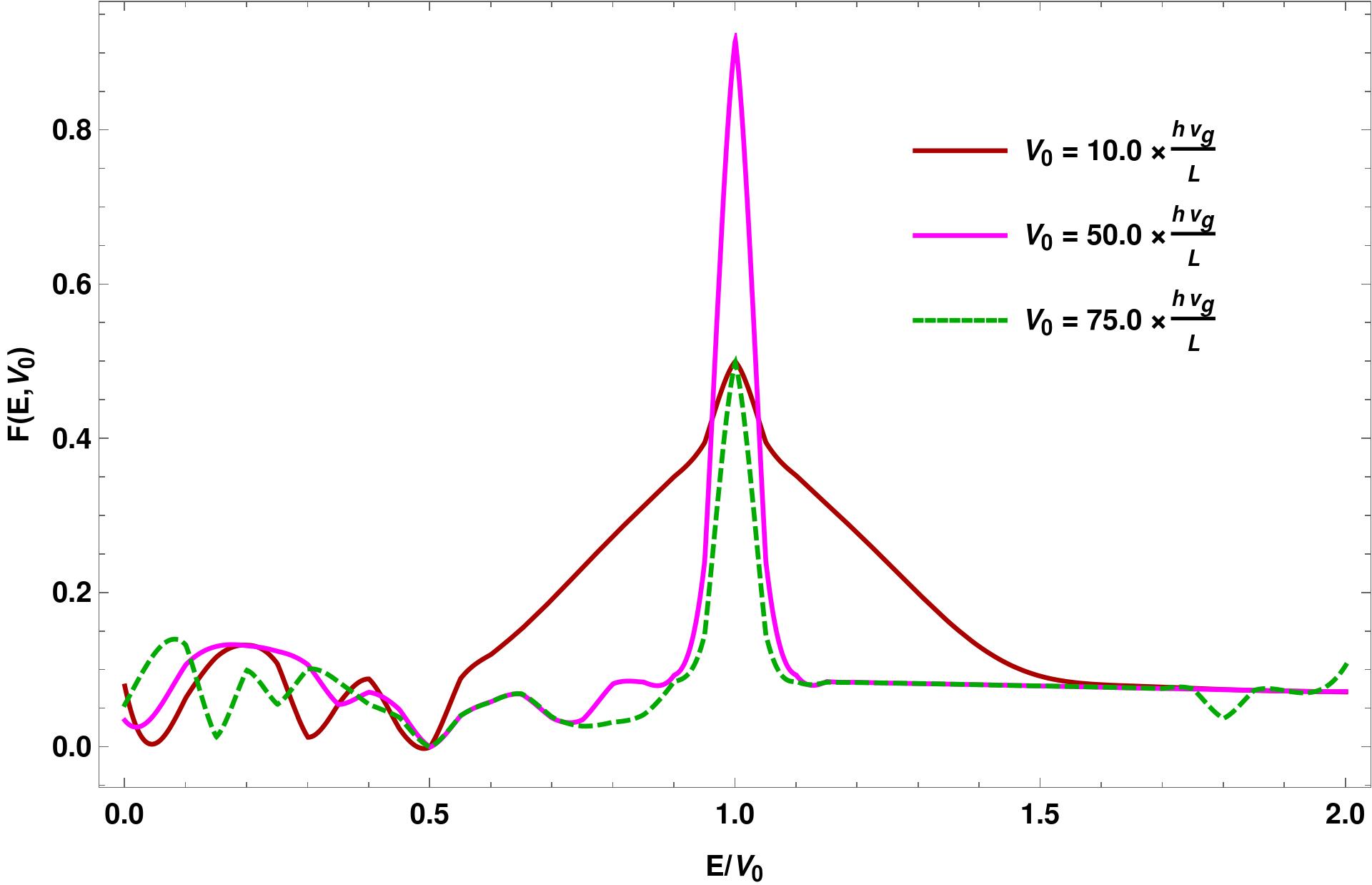}}
\caption{Pseudospin-1 semimetal: Plots of the (a) conductivity, and (b) Fano factor ($F$), as functions of $E/V_0$, for various values of $V_0$, in absence of the vector potential. $F$ is zero at $E=V_0/2$
due to super-Klein tunneling.}
\label{figfanospin1}
\end{figure}

\begin{figure}[htb]
\subfigure[]
{\includegraphics[width = 0.23 \textwidth]{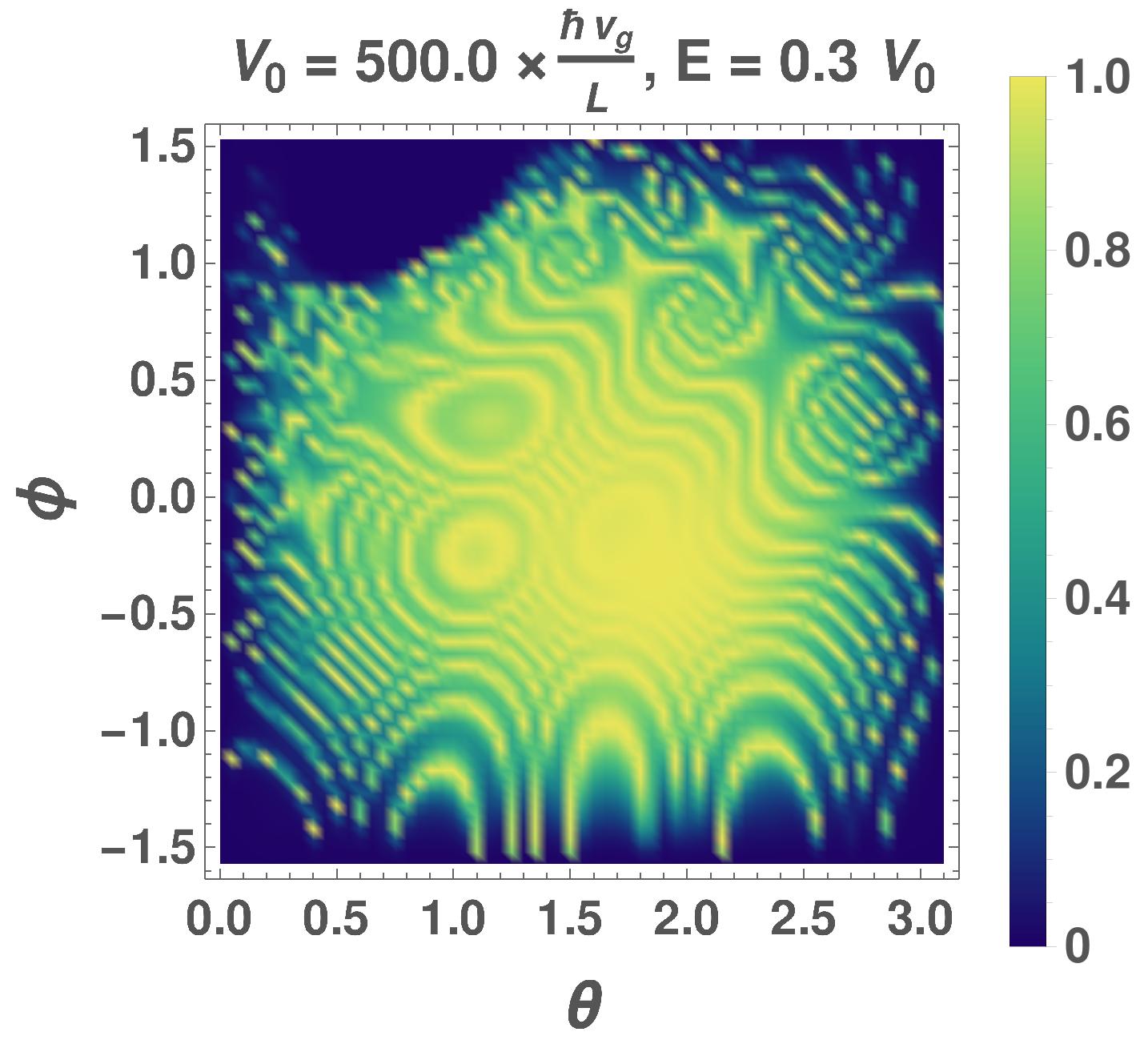}} \quad
\subfigure[]
{\includegraphics[width = 0.23 \textwidth]{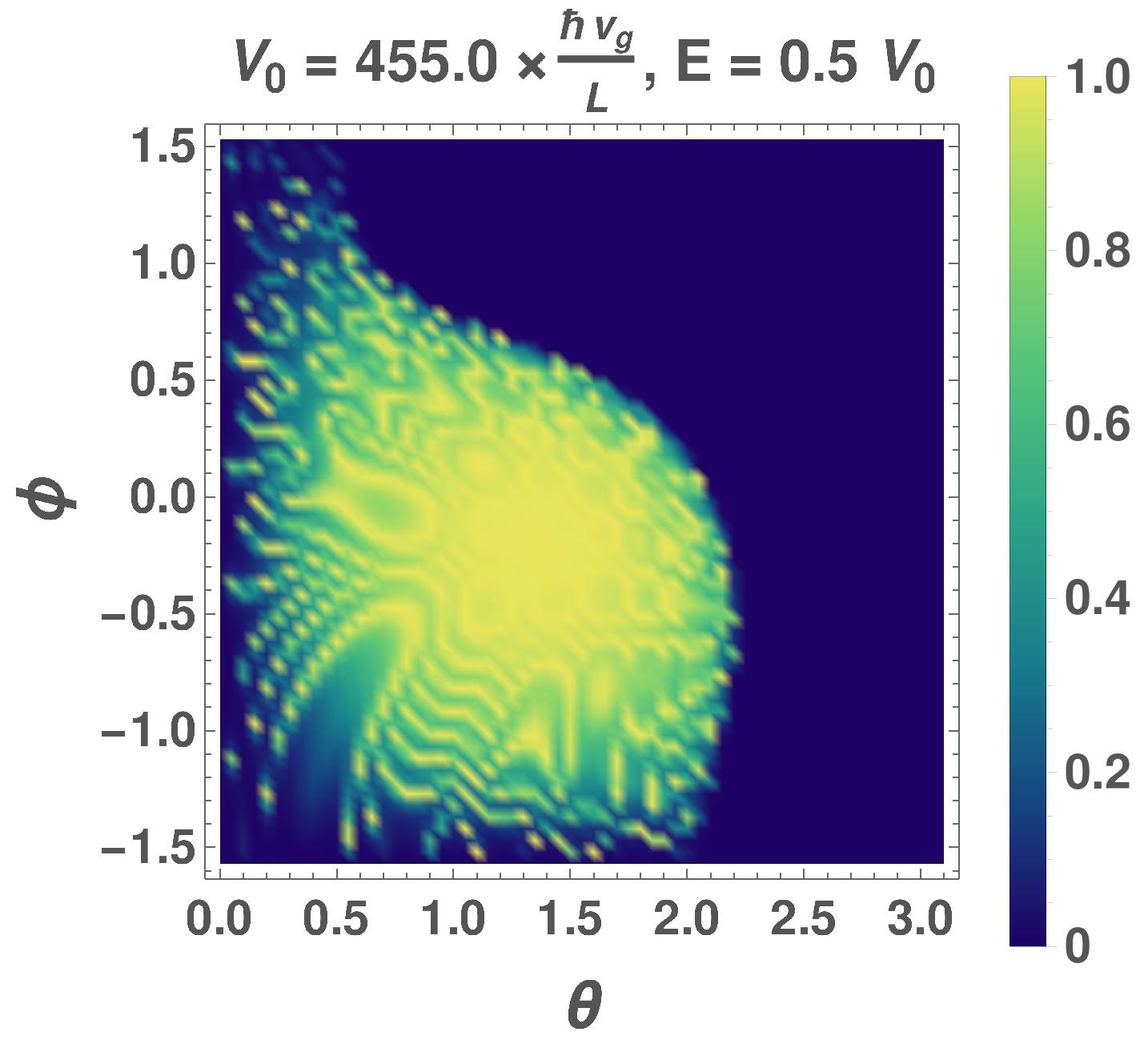}} \quad
\subfigure[]
{\includegraphics[width = 0.23 \textwidth]{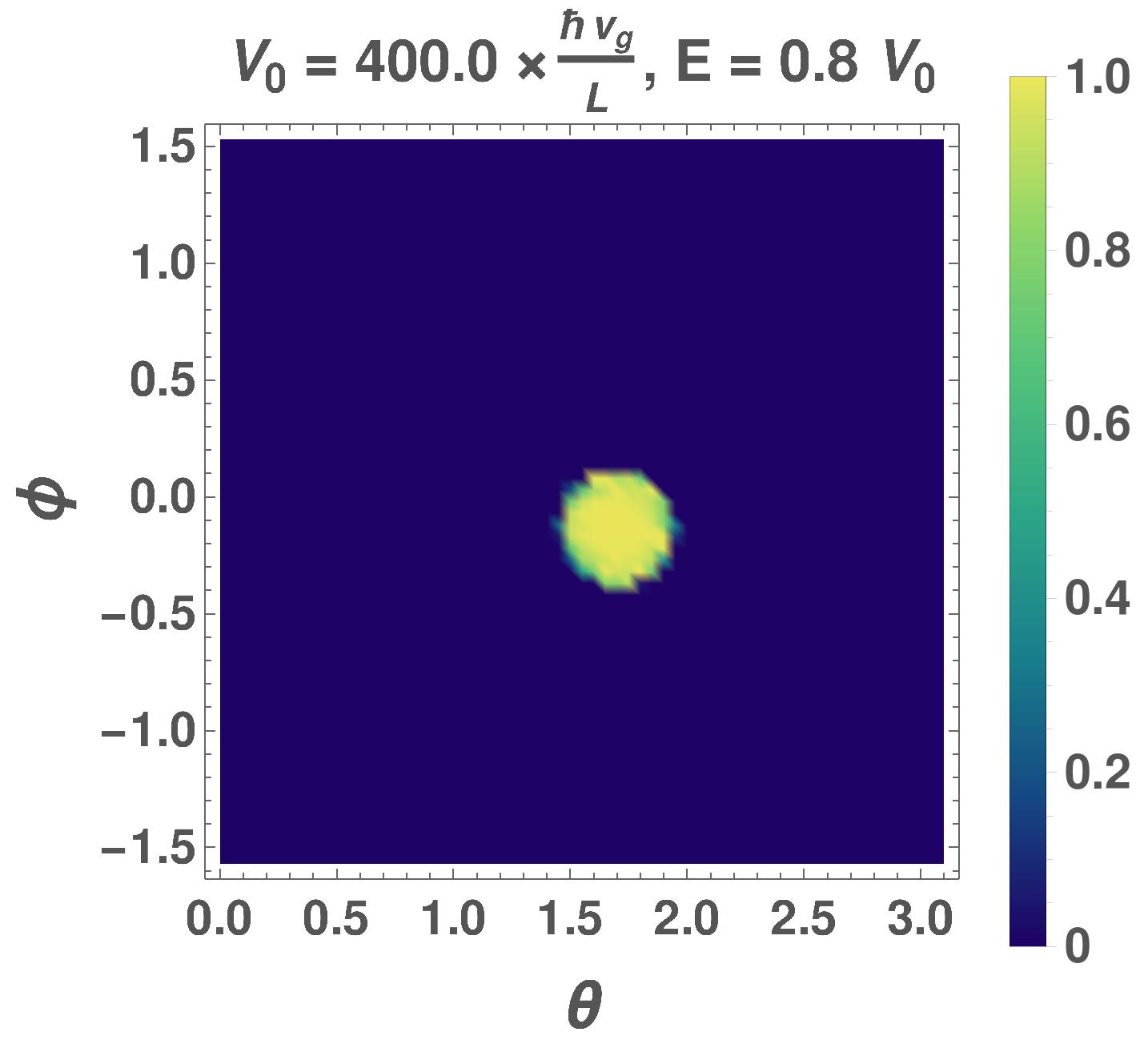}}\quad
\subfigure[]
{\includegraphics[width = 0.23 \textwidth]{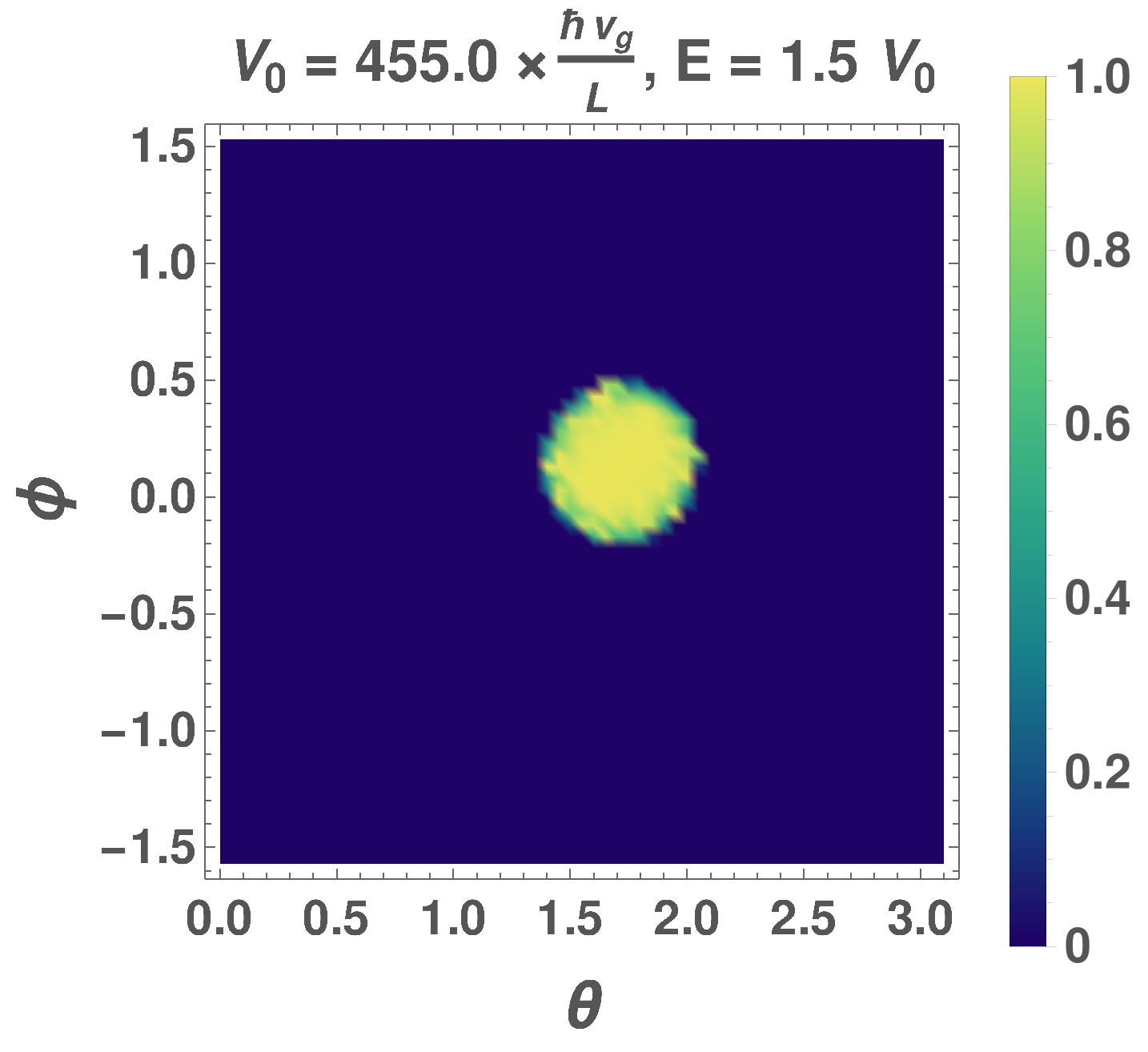}} 
\caption{Pseudospin-1 semimetal: Contourplots of the transmission coefficient ($T$) in the presence of the vector potential, as a function of $(\theta, \phi)$, for various values of $V_0$ and $E$. The values for the vector potential components $\left \lbrace  A_y, A_z \right \rbrace $ are equal to: (a) $\left \lbrace 0.5,\,0.5\right  \rbrace  
\frac{ V_0} {e\,v_g}$,
(b) $\left \lbrace 0.2,\,-0.2\right  \rbrace \frac{ V_0} {e\,v_g}$,
(c) $\left \lbrace 0.1,\,0.1\right  \rbrace \frac{ V_0} {e\,v_g}$,
(d) $\left \lbrace -0.2,\,0.2\right  \rbrace  \frac{ V_0} {e\,v_g}$.
}
\label{figcontourBspin1}
\end{figure}

We assume $ W $ to be large enough such that $\mathbf{k}_{\perp}$ can effectively be treated as a continuous variable, and perform the integrations over the angular variables to obtain conductivity and Fano factor. We express $E$ and $V_0$ in units of $ \frac{h\,v_g} { L}$.

Using $ k_\ell = \frac{E} {\hbar\,v_g} \sin \theta \cos \phi$,
$ n_y= \frac{W\, E } { h\,v_g}   \sin \theta \sin \phi ,$
$ n_z= \frac{W\, E } { h\,v_g}   \cos \theta ,$
and $ dn_y \, dn_z =
\frac{W^2 E^2 } {h^2 v_g^2}  \cos \phi \,\sin^2 \theta \, d\phi\,d\theta $, 
in the zero-temperature limit  and  for  a  small  applied  voltage,  the conductance is given by \cite{blanter-buttiker}:
\begin{align}
G(E,V_0) & = \frac{ e^2}{h} \sum_{\mathbf n}   |t_ {\mathbf n}|^2 
\rightarrow \frac{  e^2} {h} \int   |t_ {\mathbf n}|^2 \,dn_x\,dn_y
= \frac{  e^2\,W^2\, E^2 } {h^3 v_g^2} 
  \int_{ \theta=0 }^{\pi} \int_{\phi=-\frac{\pi}{2}}^{\frac{\pi}{2}}
T( E ,V_0,\theta, \phi,\mathbf B) \cos\phi \sin^2 \theta \,d\phi \, d \theta,
\end{align}
leading to the conductivity expression:
\begin{align}
\sigma (E,V_0,\mathbf B)  & = \left( \frac{L }{W} \right)^2 \frac{G(E,V_0)}{e^2/h}
 =   \left( \frac{ E } {h v_g/L} \right)^2
\int_{ \theta=0 }^{\pi} \int_{\phi=-\frac{\pi}{2}}^{\frac{\pi}{2}}
T( E ,V_0,\theta, \phi,\mathbf B) \cos\phi \sin^2 \theta \,d\phi\,d\theta.
\end{align}
The Fano factor can be expressed as:
\begin{align}
F(E,V_0,\mathbf B)  &=\frac 
{\int_{ \theta=0 }^{\pi} \int_{\phi=-\frac{\pi}{2}}^{\frac{\pi}{2}}
T( E ,V_0,\theta, \phi,\mathbf B) 
 \left[1-T( E ,  V_0, \theta, \phi,\mathbf B) \right]
\cos\phi \sin^2 \theta \,d\phi \,d\theta  } 
{ \int_{ \theta=0 }^{\pi} \int_{\phi=-\frac{\pi}{2}}^{\frac{\pi}{2}}
T( E ,V_0,\theta, \phi,\mathbf B) \cos\phi \sin^2 \theta \,d\phi\,d\theta } \,.
\end{align}

First let us study the characteristics of transmission coefficients in the absence of the magnetic fields. 
Fig.~\ref{figpolarspin1} shows the polar plots of $T( E , V_0,\pi/2,\phi,\mathbf 0)$ as a function of the incident angle $\phi$ (at $\theta=\pi/2$), which corresponds to $k_z=0$.
In Fig.~\ref{figcontourspin1}, we show the angular dependence of $T( E , V_0,\theta,\phi,\mathbf 0)$ in contourplots. As $E$ approaches the value $V_0/2$, it reaches the condition of super-Klein tunneling where there is perfect transmission for all angles. The super-Klein contourplot is not shown here as this would have been a redundant plot. As $E$ goes above $V_0/2$, the transmission regions get confined to narrower and narrower angular regions, centred around $(\theta=0, \phi=0)$. 
In Fig.~\ref{figfanospin1}, we illustrate the conductivity $\sigma (E,V_0,\mathbf 0)$ and the Fano factor $ F (E,V_0,\mathbf 0)$, as functions of $E/V_0$, for some values of $V_0$. Due to super-Klein tunneling,
$F=0$ for $E =V_0/2$.

The presence of the vector potential modifies the contourplots of $T$, as shown in Fig.~\ref{figcontourBspin1}. Although $T=1$ for $E =V_0/2$ (for all angles) in absence of magnetic fields, this feature is destroyed by the constant vector potential, as seen in Fig.~\ref{figcontourBspin1}(b).
For values of $E$ above $V_0/2$, the $T \simeq 1$ regions get restricted to discs (just like in the $\mathbf{B}=0$ case), whose centres are now shifted away from the $(\theta=\pi/2, \phi=0)$ point due to the effect of $\mathbf{B}\neq 0$.

\section{Pseudospin-3/2 Fermions}
\label{secspin32}

The eight space groups $207$-$214$ can host fourfold topological degeneracies about the $\Gamma $, R and/or H points \cite{bernavig}. The linearized $\mathbf{k} \cdot \mathbf {p}$ Hamiltonian about such a point hosts pseudospin-3/2 fermions and takes the form:
\begin{align}
\mathcal{H}_{3/2}(\mathbf  k) = \hbar\,v_g \,\mathbf{k}\cdot \mathbf J\,,
\end{align}
where the three components of $\mathbf J$ form the spin-3/2 representation of the SO(3) group, and their standard representation is given by:
\begin{align}
J_x = \left(
\begin{array}{cccc}
 0 & \frac{\sqrt{3}}{2} & 0 & 0 \\
 \frac{\sqrt{3}}{2} & 0 & 1 & 0 \\
 0 & 1 & 0 & \frac{\sqrt{3}}{2} \\
 0 & 0 & \frac{\sqrt{3}}{2} & 0 \\
\end{array}
\right), \, \,
J_y = \left(
\begin{array}{cccc}
 0 & \frac{-\sqrt{3} \,\mathrm{i}} {2}   & 0 & 0 \\
 \frac{ \sqrt{3}\,\mathrm{i}}{2} & 0 & - \mathrm{i} & 0 \\
 0 & \mathrm{i} & 0 & \frac{ - \sqrt{3}\,\mathrm{i}}{2}  \\
 0 & 0 & \frac{ \sqrt{3} \,\mathrm{i}} {2} & 0 \\
\end{array}
\right)\,,\,\,
J_z =\frac{1}{2} \left(
\begin{array}{cccc}
 3 & 0 & 0 & 0 \\
 0 & 1 & 0 & 0 \\
 0 & 0 & -1 & 0 \\
 0 & 0 & 0 & -3 \\
\end{array}
\right)\,.
\end{align}
Here $v_g$ denotes the magnitude of the group velocity of the quasiparticles. 
The energy eigenvalues take the form:
\begin{align}
\varepsilon_{3/2}^{\pm}(\mathbf{k}) = \pm \frac{ 3\,\hbar \, v_g  k} {2} \,,
\quad \varepsilon_{1/2}^{\pm}(\mathbf{k}) = \pm \frac{ \hbar \, v_g  k} {2} \,,
\end{align}
demonstrating four linearly dispersing bands crossing at a point.
Here the ``$+$" and ``$-$" signs, as usual, refer to the conduction and valence bands, respectively.
The corresponding normalized eigenvectors are given by:
\begin{align}
\Psi^s_{3/2} & = \frac{1} {\mathcal{N}^s_{3/2} }
\left\{\frac{ s\,k \left( k_x^2+k_y^2+4\, k_z^2\right)+
k_z \left(3\, k_x^2+3 k_y^2+4 \,k_z^2\right)} { \left (k_x+  \mathrm{i} \, k_y \right )^3},
\frac{\sqrt{3} \left [ 2\, k_z \left (s\, k+k_z \right )+k_x^2+k_y^2\right ]}
{\left (k_x+ \mathrm{i} \,k_y\right )^2},\frac{\sqrt{3} \left (s\, k+k_z \right )}
{k_x+ \mathrm{i}\,k_y},1\right\}^T,\nn
\Psi^s_{1/2} & = \frac{1} { \mathcal{N}^s_{1/2} }
\left\{-\frac{ \left( s\, k+k_z  \right )  \left (k_x- \mathrm{i}\, k_y \right )}
{(k_x+ \mathrm{i}\,k_y)^2},-\frac{-2 \,k_z \left ( s\, k+k_z \right )+k_x^2+k_y^2}
{\sqrt{3} \left (k_x+ \mathrm{i}\, k_y \right )^2},\frac{ s\, k+3 \, k_z}{\sqrt{3} 
\left (k_x+ \mathrm{i}\, k_y \right )},1\right\}^T\,,
\end{align}
respectively, where  $s=\pm $, and $ \frac{1} {\mathcal{N}^s_{3/2} }$ and $\frac{1} {\mathcal{N}^s_{1/2}}$ denote the corresponding normalization factors.

In presence of the scalar and vector potentials, we need to consider the total Hamiltonian:
\begin{align}
\mathcal{H}_{3/2}^{tot} &=
 \mathcal{H}_{3/2}
\left [-\mathrm{i}\,\nabla + \frac{e\,\mathbf A(x)}{\hbar} \right ]
 +V(x)
\end{align} 
in position space, and find the appropriate wavefunctions.

\subsection{Mode-matching}

We will follow the same procedure as described for the pseudospin-1 semimetals. Again, without any loss of generality, we consider the transport of one of the positive energy states, namely $\Psi^+_{3/2} $, corresponding to electron-like particles, with the Fermi level outside the potential barrier being adjusted to the value $E = \frac{ 3\,\hbar \, v_g  k} {2} $.
In this case, a scattering state $\tilde \Psi_{ \mathbf n}$, in the mode labeled by $\mathbf n$, is
constructed from the states:
\begin{align}
 \tilde \Psi_{ \mathbf n} (x)=&  \begin{cases} \tilde \phi_L & \text{ for } x<0   \\
\tilde \phi_M & \text{ for } 0< x < L \\
\tilde \phi_R &  \text{ for } x > L 
\end{cases} \,,
\end{align}
where
\begin{align}
 \tilde \phi_L = & \,  \frac{
 \Psi^+_{3/2} (   k_{\ell,3/2}\, ,  \tilde {\mathbf{k}}_\perp) \, e^{\mathrm{i}\, k_{\ell,3/2}\, x }}
{\sqrt{ \tilde{ \mathcal{V} }(k_{\ell, 3/2}\,,\, \mathbf n) }} 
+  
\sum \limits_{\sigma = \frac{1}{2},  \frac{3}{2}}
 \frac{  r_{\mathbf n,\sigma} \,\Psi^+_{\sigma} ( -k_{\ell,\sigma} , \tilde {\mathbf{k}}_\perp) \, e^{-\mathrm{i}\, k_{\ell,\sigma}\, x }
}
{\sqrt{ \tilde{ \mathcal{V} }(k_{\ell,\sigma}, \mathbf n) }}
\, ,\nonumber \\
 \tilde \phi_M  = &\,\Big[  
 \sum \limits_{\sigma = \frac{1}{2},  \frac{3}{2}}
 \alpha_{\mathbf n,\sigma} \,\Psi^{+}_\sigma ( \tilde k_\sigma,  \tilde {\mathbf{k}}_\perp) \,
 e^{\mathrm{i}\,\tilde k_\sigma\,  x } 
 + 
\sum \limits_{\sigma = \frac{1}{2},  \frac{3}{2}}
 \beta_{\mathbf n,\sigma } \,\Psi_\sigma^{+} (  -\tilde k_\sigma,  \tilde {\mathbf{k}}_\perp) \,
 e^{-\mathrm{i}\,\tilde k_\sigma  \, x } 
\Big]   \Theta\left( E-V_0  \right)
  \nonumber\\
& 
 + \Big[  
 \sum \limits_{\sigma = \frac{1}{2},  \frac{3}{2}}
 \alpha_{\mathbf n,\sigma} \,\Psi^{-}_\sigma ( \tilde k_\sigma,  \tilde {\mathbf{k}}_\perp) \,
 e^{\mathrm{i}\,\tilde k_\sigma\,  x } 
 + 
\sum \limits_{\sigma = \frac{1}{2},  \frac{3}{2}}
 \beta_{\mathbf n,\sigma } \,\Psi_\sigma^{-} (  -\tilde k_\sigma,  \tilde {\mathbf{k}}_\perp) \,
 e^{-\mathrm{i}\,\tilde k_\sigma  \, x } 
\Big]   \,\Theta\left( V_0-E  \right) \,,\nonumber \\
\tilde \phi_R = & \sum \limits_{\sigma = \frac{1}{2},  \frac{3}{2}}
 \frac{  t_{\mathbf n,\sigma} \,\Psi^+_{\sigma} ( k_{\ell,\sigma},  \tilde {\mathbf{k}}_\perp) 
\, e^{ \mathrm{i}\, k_{\ell,\sigma}\, x }
}
{\sqrt{ \tilde{ \mathcal{V} }(k_{\ell,\sigma}, \mathbf n) }},\nonumber \\
\tilde { \mathcal{V} }(k_{\ell,\sigma}, \mathbf n) \equiv &  \, |\partial_{k_\ell} \varepsilon_{\sigma}^+ (k_\ell, \mathbf n)|\,,
\quad k_{\ell,3/2} = \sqrt{\frac{ 4\,E^2} {9\, \hbar^2 v_g^2}-  \mathbf{k}_\perp^2}
\,,\quad  
\tilde k_{3/2} = \sqrt{\frac{ 4\left( E-V_0\right)^ 2} { 9\,\hbar^2 v_g^2} 
-{ \tilde {\mathbf{k}}_\perp} ^2 }\,,\quad 
 \tilde {\mathbf{k}}_\perp = \mathbf{k}_\perp +  \frac{e\,\mathbf A(x)}{\hbar} \,,
\nn k_{\ell,1/2} =& \sqrt{\frac{ 4\,E^2} {\hbar^2 v_g^2}-  \mathbf{k}_\perp^2}
\,,\quad  
\tilde k_{1/2} = \sqrt{\frac{ 4\left( E-V_0\right)^ 2} { \hbar^2 v_g^2} 
-{ \tilde {\mathbf{k}}_\perp} ^2 }\,.
\end{align}
We have used the velocity $ \tilde{ \mathcal{V} }(k_{\ell,\sigma}, \mathbf n) $ to normalize the incident, reflected and transmitted plane waves. 

The usual mode-matching procedure at $x=0$ and $x=L$ allows us to solve for
$t_{\mathbf n,\sigma}(E, V_0,\mathbf B) $ numerically.
The transmission probabilities at an energy $E$ are given by: 
\begin{align}
T_\sigma( E ,  V_0,\theta , \phi,\mathbf B) = | t_{\mathbf n,\sigma}( E, V_0,\mathbf B )|^2 \,,
\text{ where } 
\theta = \cos^{-1} \left( \frac{3\, \hbar\,v_g \,q_{n_z}} { 2\,E } \right)
\text{ and }
\phi = \tan^{-1} \left( \frac {q_{n_y}} {k_{\ell,3/2} } \right)
\end{align}
define the incident angle (solid) of the incoming wave in 3d.
For normal incidence, we get the simple analytical expression $t_{\mathbf 0,\sigma}(E, V_0,\mathbf 0) = e^{\frac{\mathrm{i}\, L \left( E-V_0\right) }{3} }\,\delta_{\sigma,3/2}\,$, which implies the occurrence of Klein tunneling with perfect transmission ($T_{3/2} =1 $ and $T_{1/2} =0 $).
We note that super-Klein tunneling \cite{lai,zhu} is absent for the pseudospin-3/2 quasiparticles, unlike
the pseudospin-1 Dirac cone systems.

\begin{figure}[h!]
\subfigure[]{\includegraphics[width = 0.14 \textwidth]{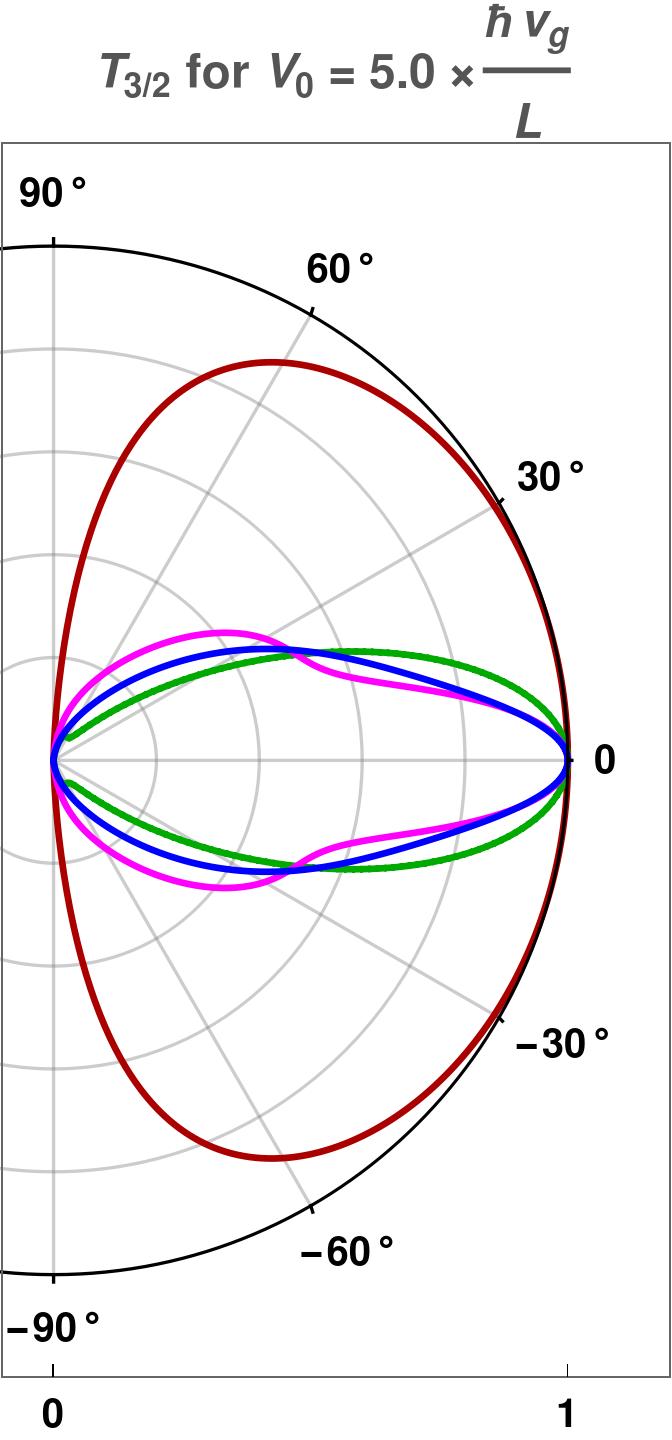}} \quad
\subfigure[]{\includegraphics[width = 0.14 \textwidth]{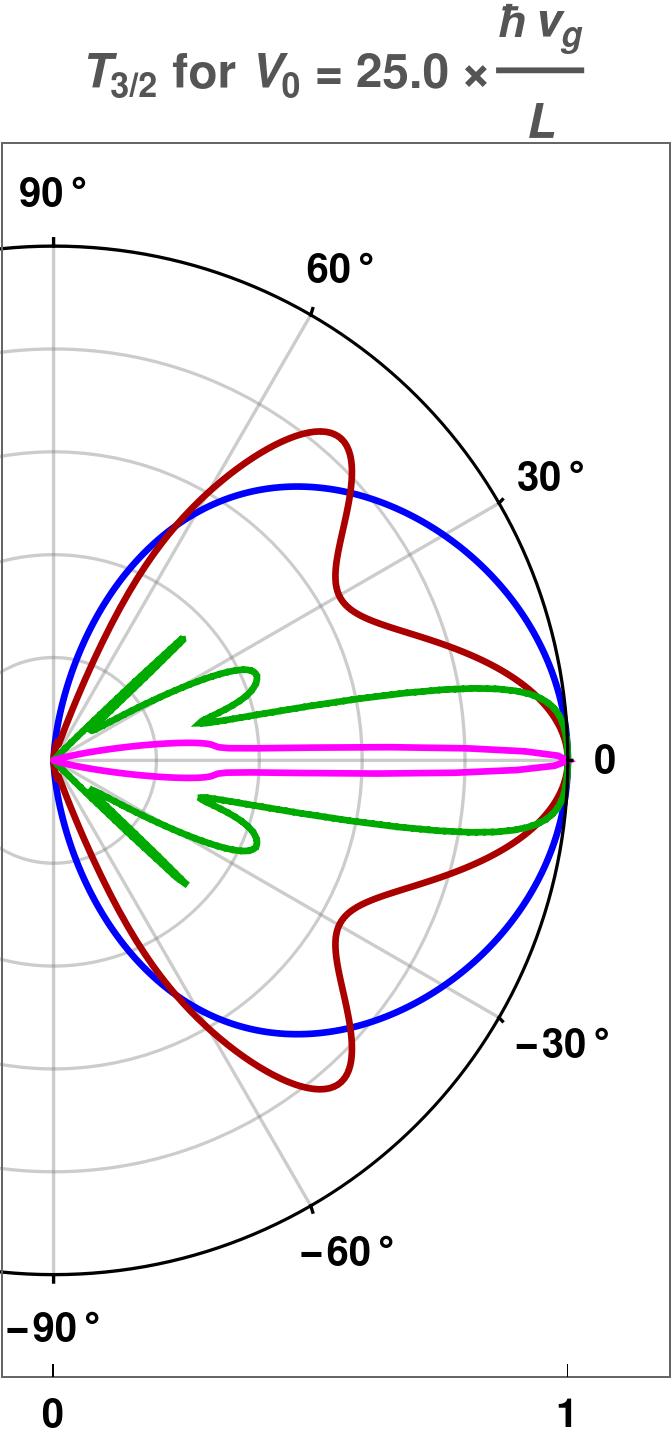}}\quad
\subfigure[]{\includegraphics[width = 0.14 \textwidth]{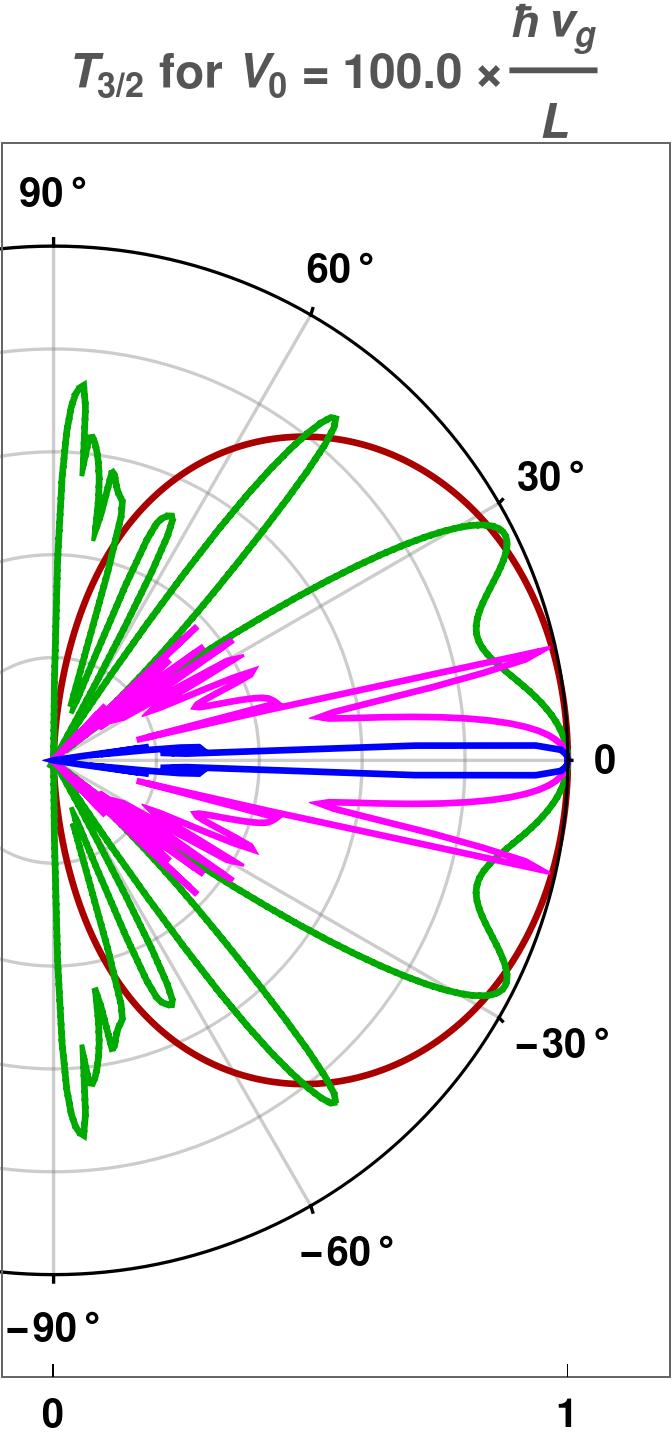}}  \quad
\subfigure[]{\includegraphics[width = 0.14 \textwidth]{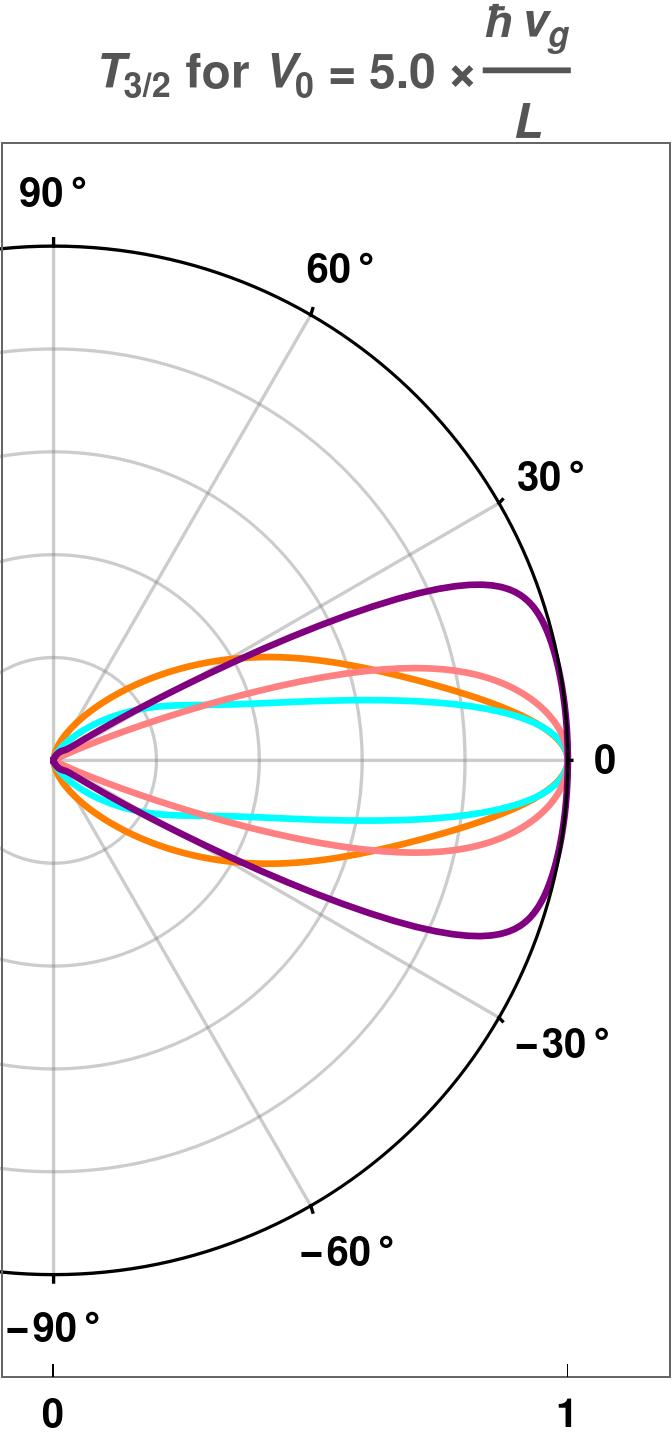}}  \quad
\subfigure[]{\includegraphics[width = 0.14 \textwidth]{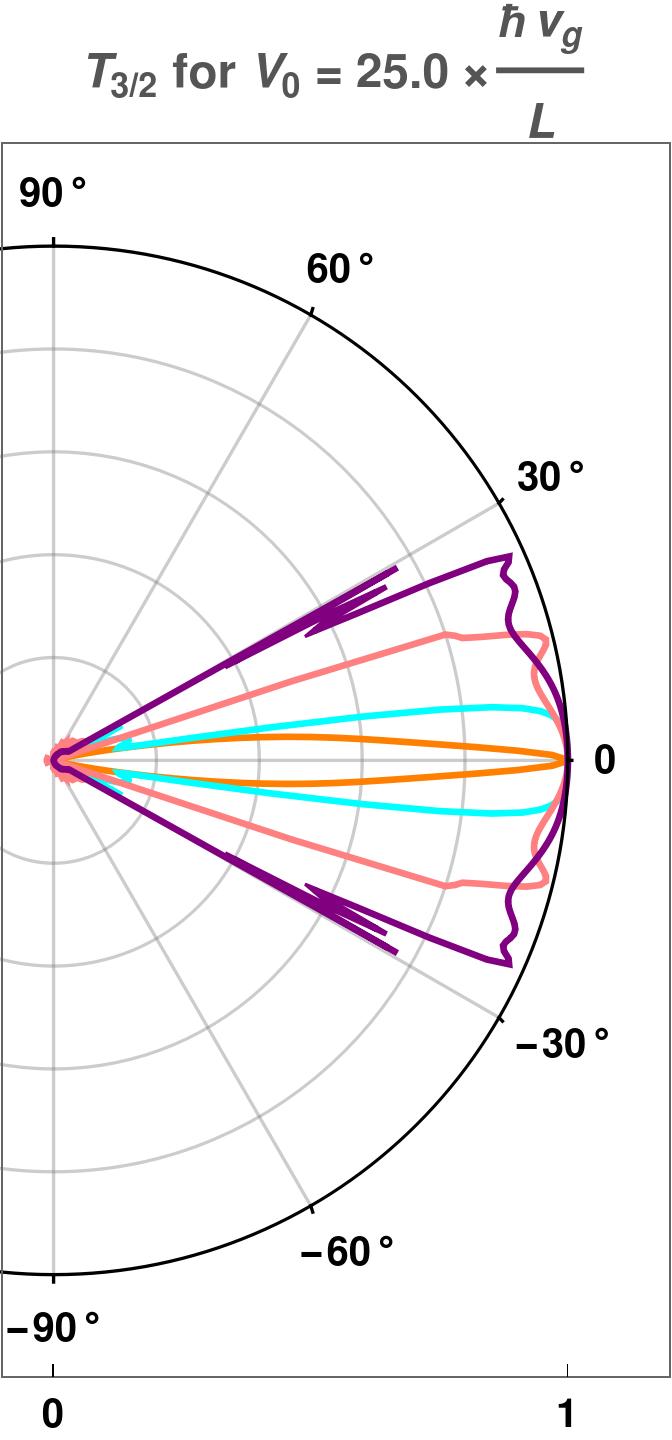}}  \quad
\subfigure[]{\includegraphics[width = 0.14 \textwidth]{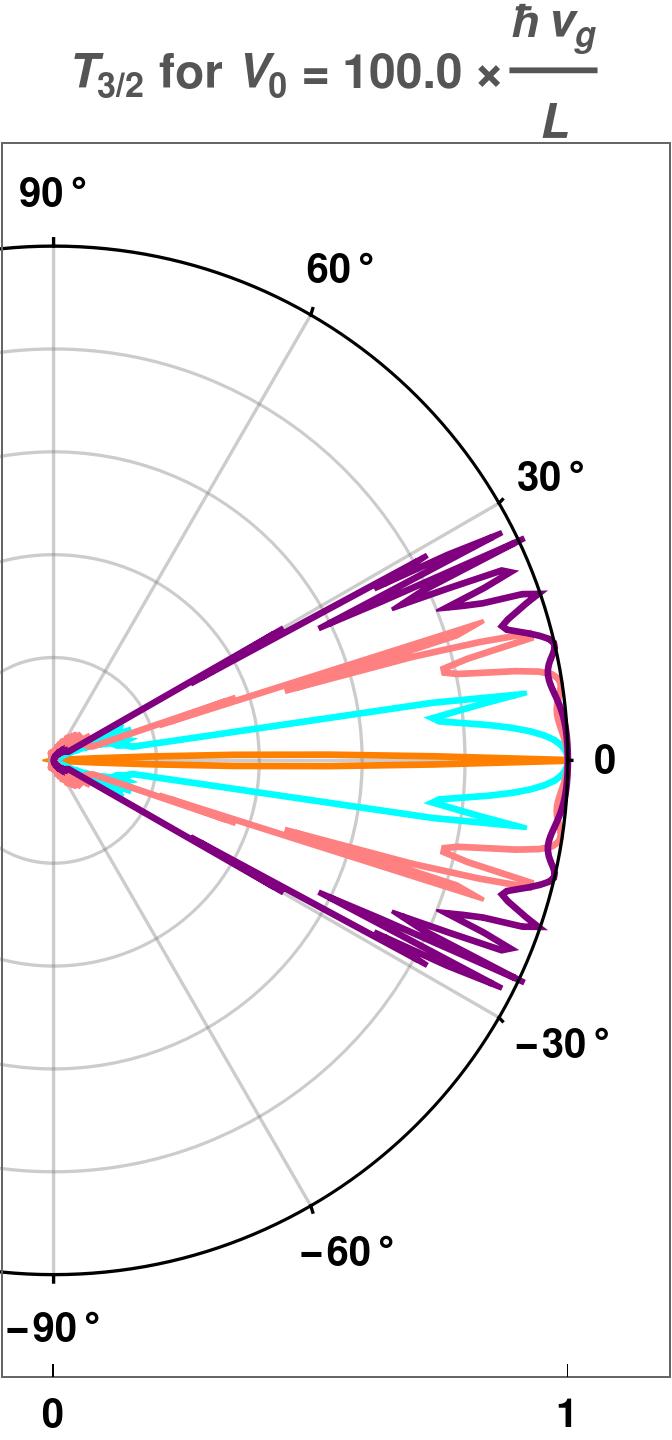}}
\subfigure[]{\includegraphics[width = 0.14 \textwidth]{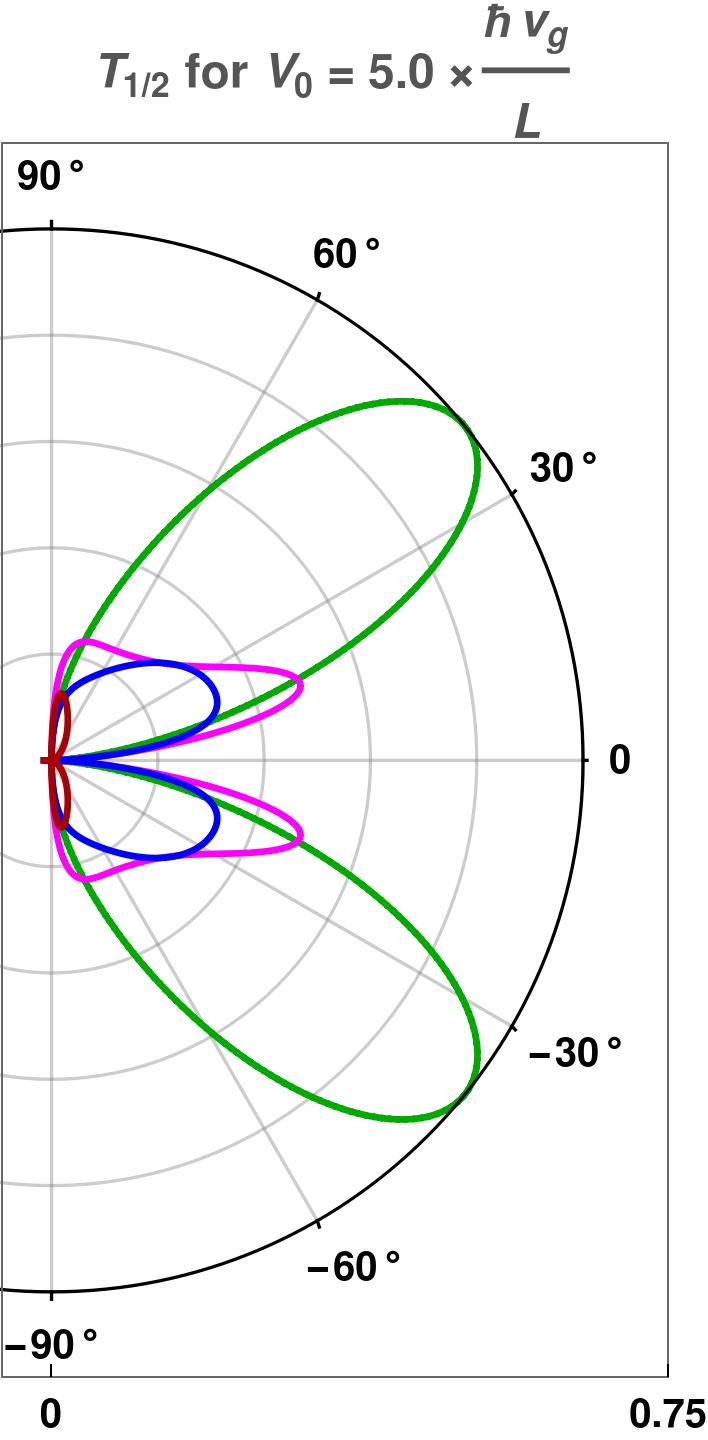}} \quad
\subfigure[]{\includegraphics[width = 0.14 \textwidth]{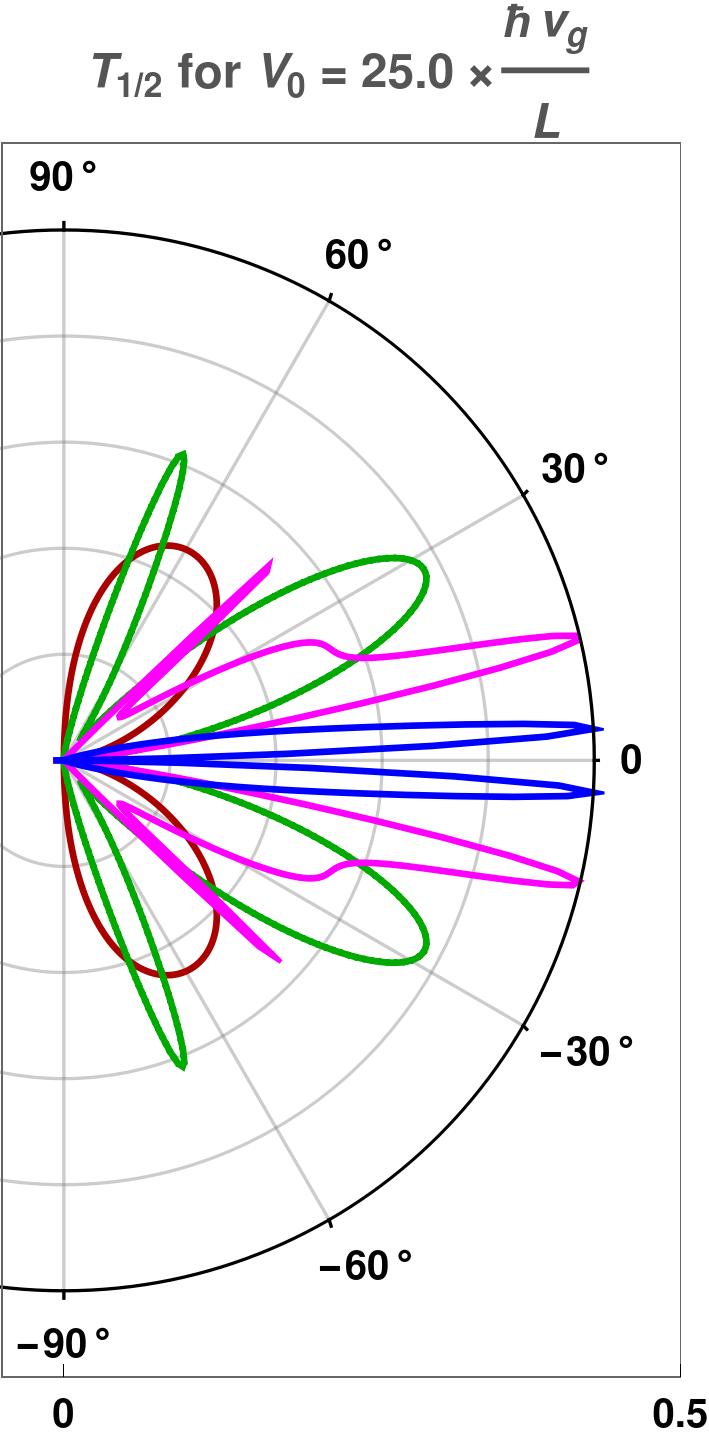}}\quad
\subfigure[]{\includegraphics[width = 0.128 \textwidth]{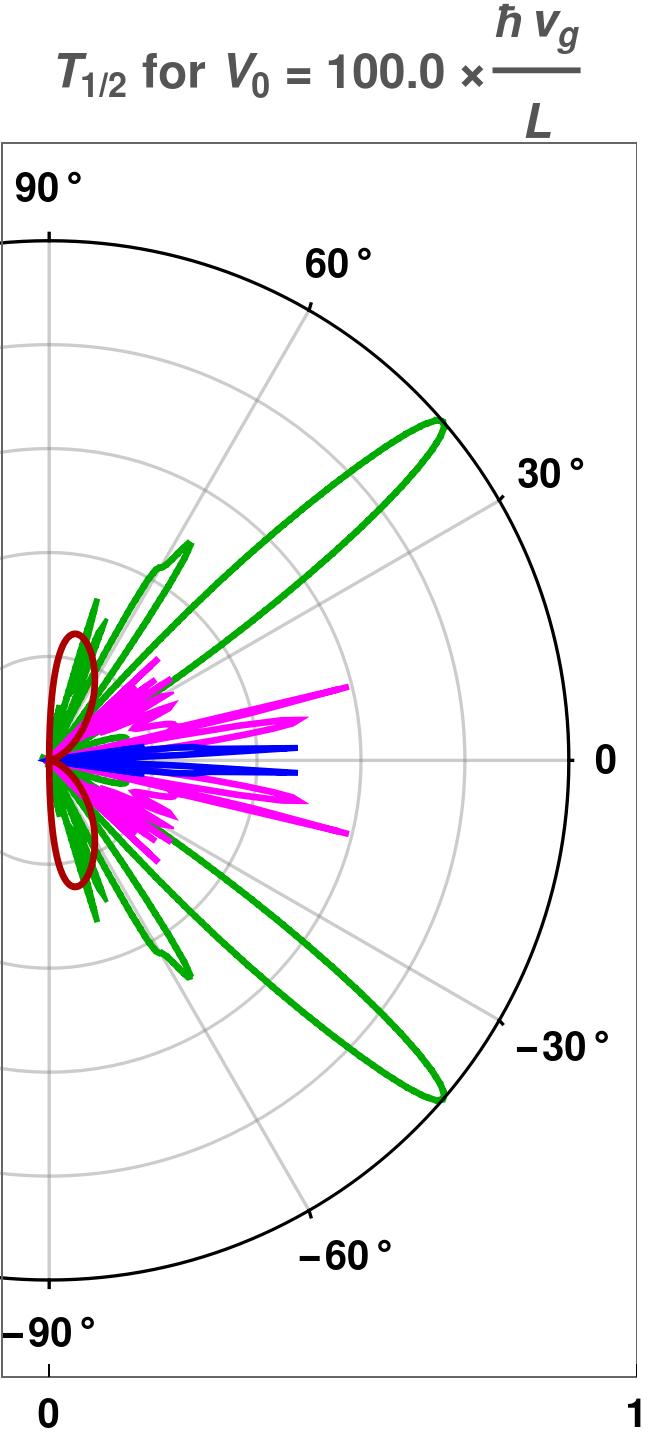}}  \quad
\subfigure[]{\includegraphics[width = 0.14 \textwidth]{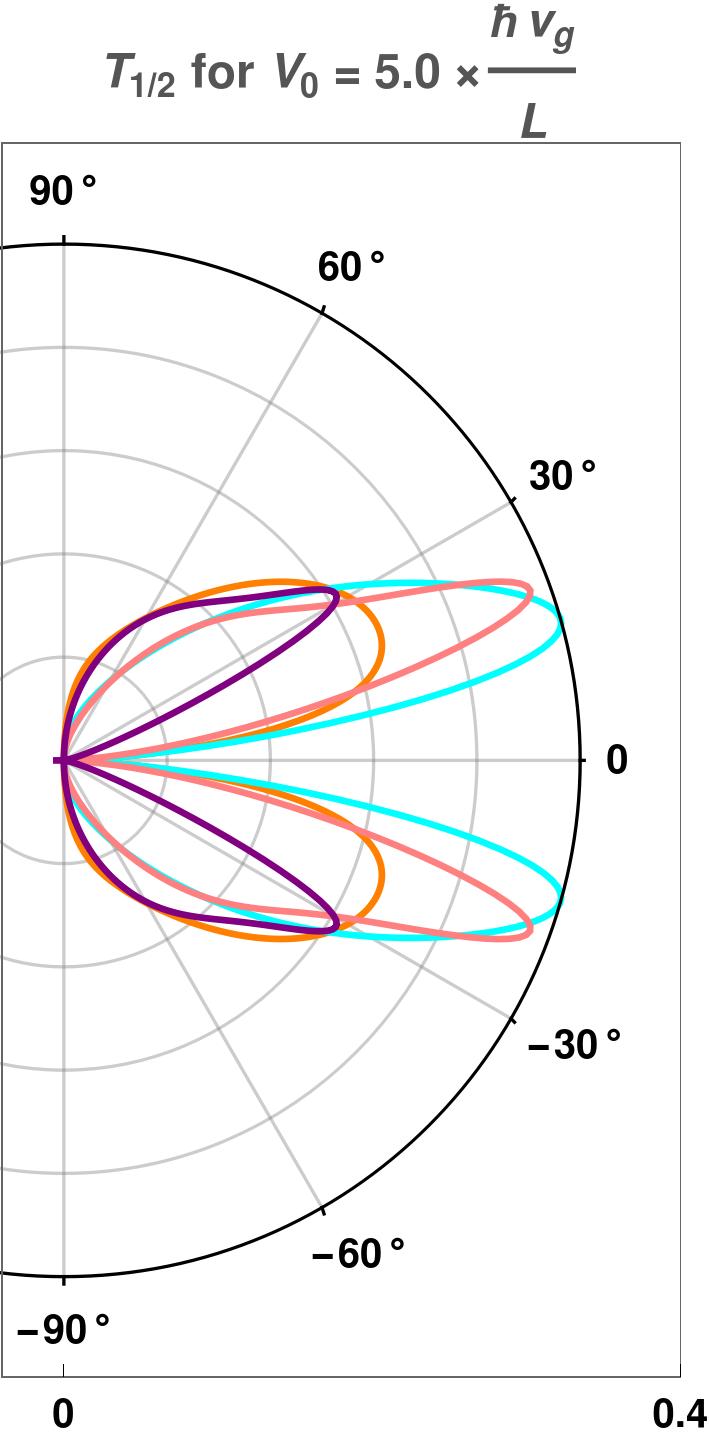}}  \quad
\subfigure[]{\includegraphics[width = 0.14 \textwidth]{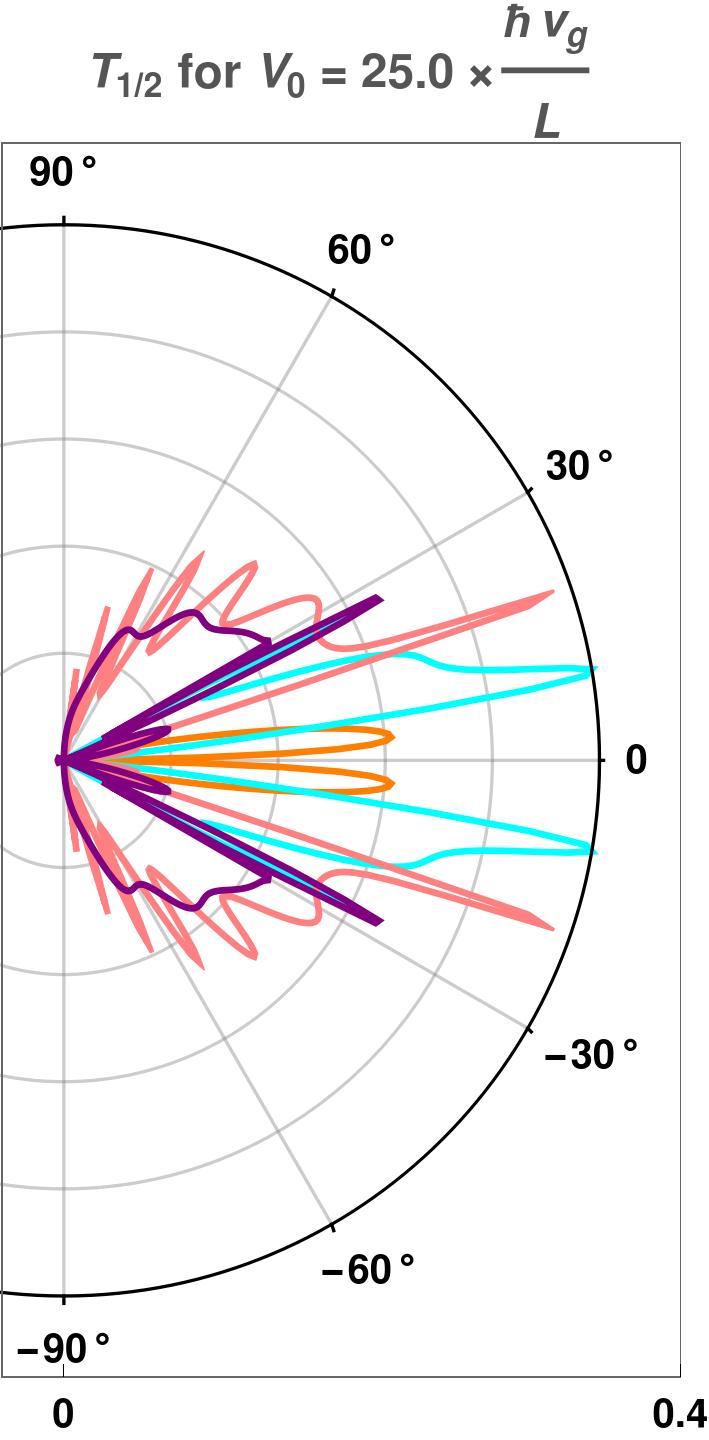}}  \quad
\subfigure[]{\includegraphics[width = 0.138 \textwidth]{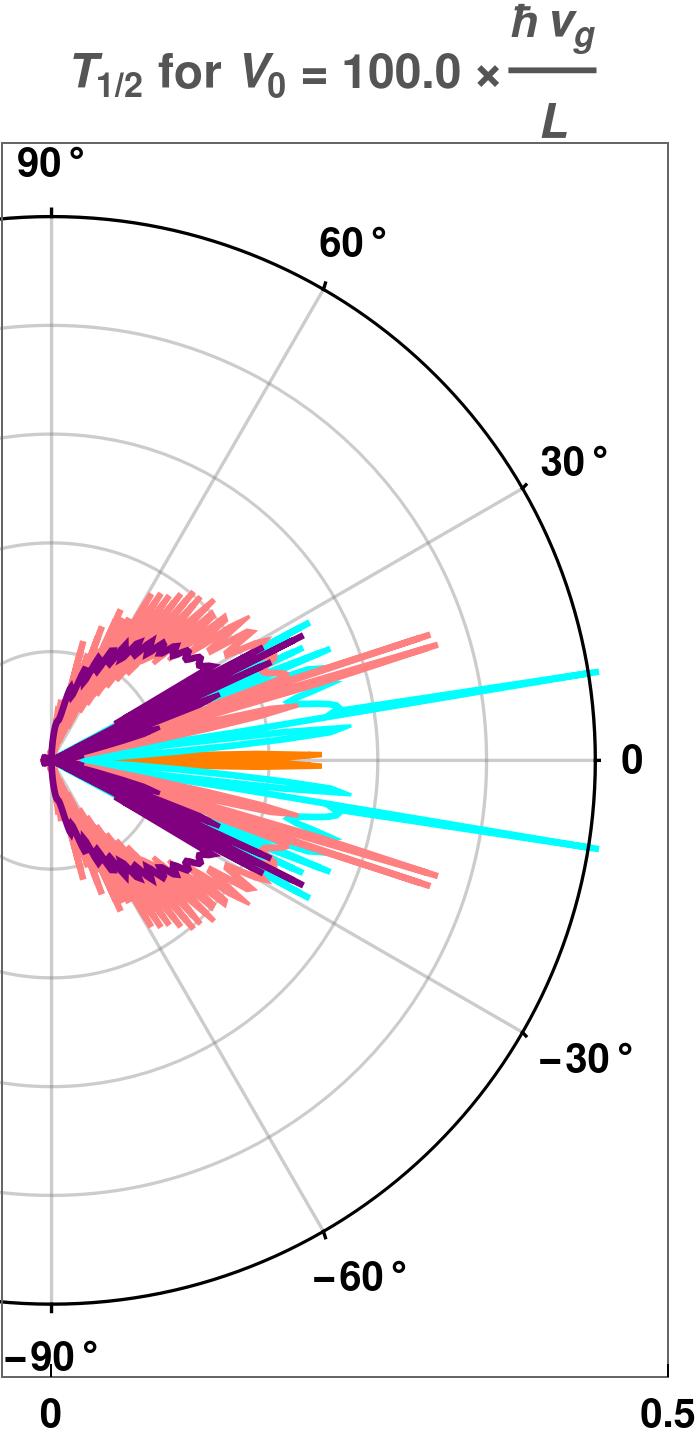}}
\caption{Pseudospin-3/2 semimetal: The polar plots show the transmission coefficients $T_\sigma(E, V_0,\theta,\frac{\pi}{2},\mathbf 0) $ as functions of the incident angle $\phi$ (in the $xy-$plane with no $k_z-$component) for the parameters $E= 0.3 \,V_0$ (red), $E=0.5\,V_0$ (green), $E=0.8\,V_0$ (magenta), $E=0.95\,V_0$ (blue), $E=1.001\,V_0$ (orange), $E=1.2 \,V_0$ (cyan), $E=1.5\,V_0$ (pink), and $E=2.0 \,V_0$ (purple). Klein tunneling is observed for normal incidence.}
\label{figpolarspin32}
\end{figure}

\begin{figure}[h!]
\subfigure[]
{\includegraphics[width = 0.23 \textwidth]{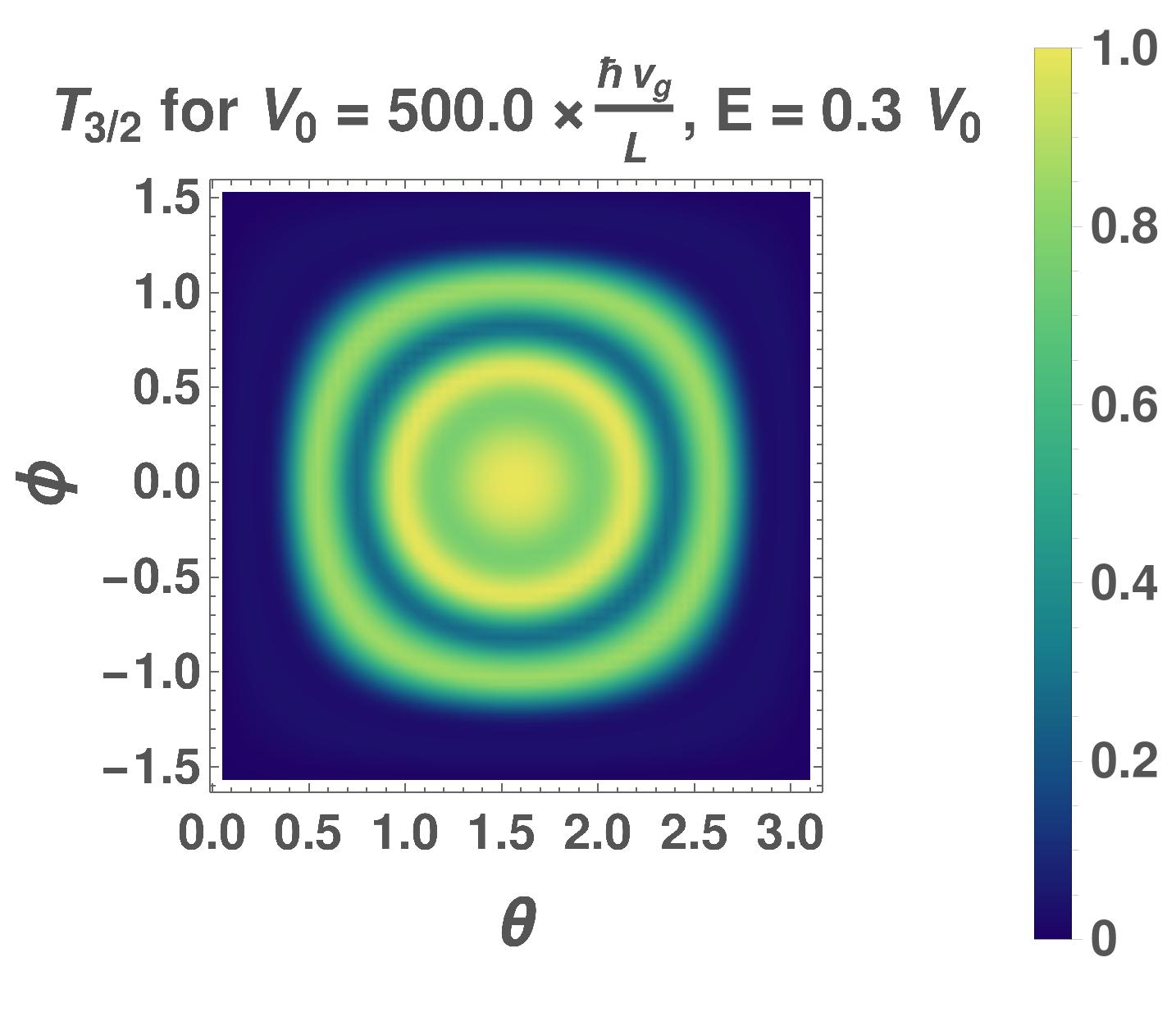}} \quad
\subfigure[]
{\includegraphics[width = 0.23 \textwidth]{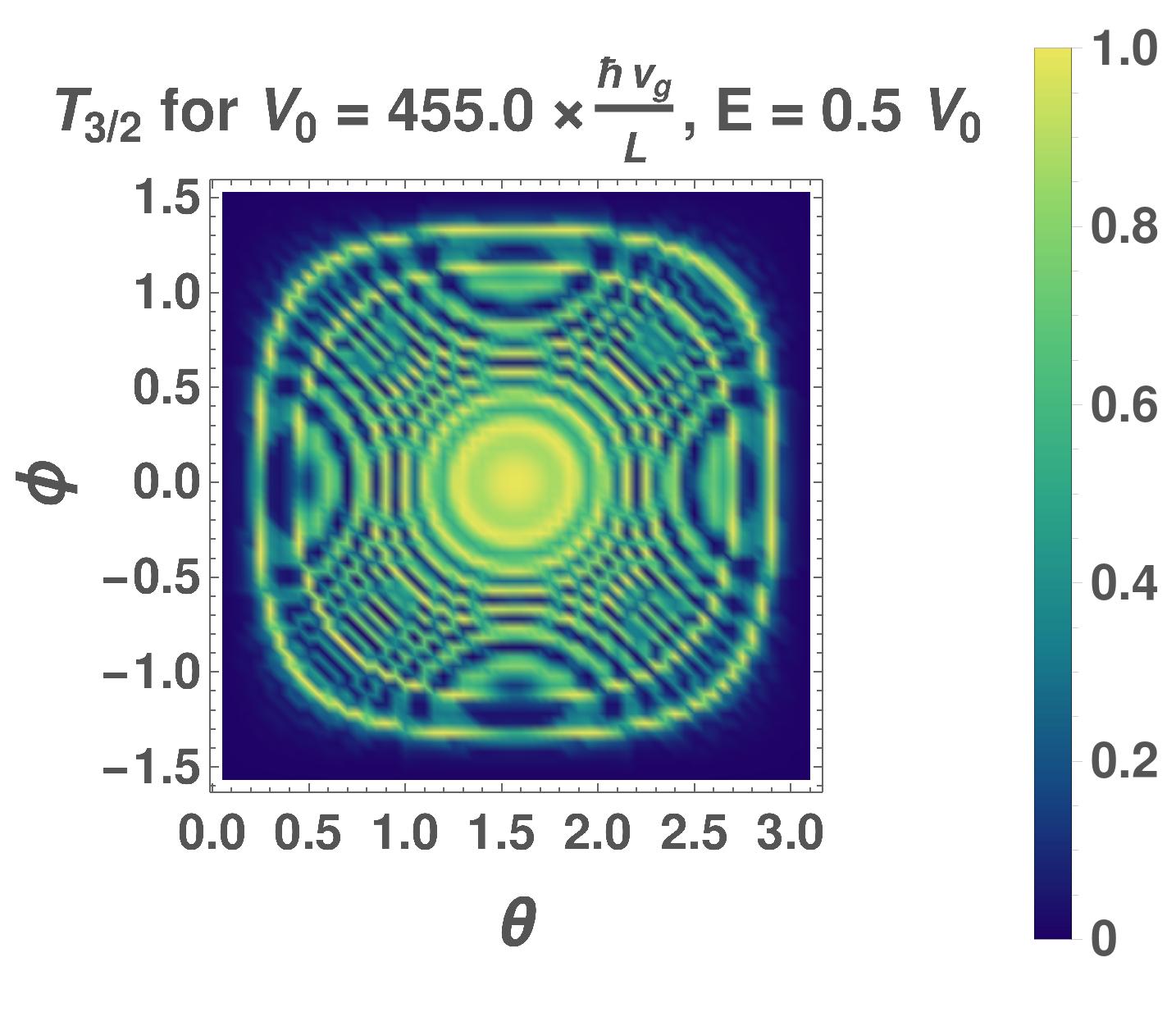}} \quad
\subfigure[]
{\includegraphics[width = 0.23 \textwidth]{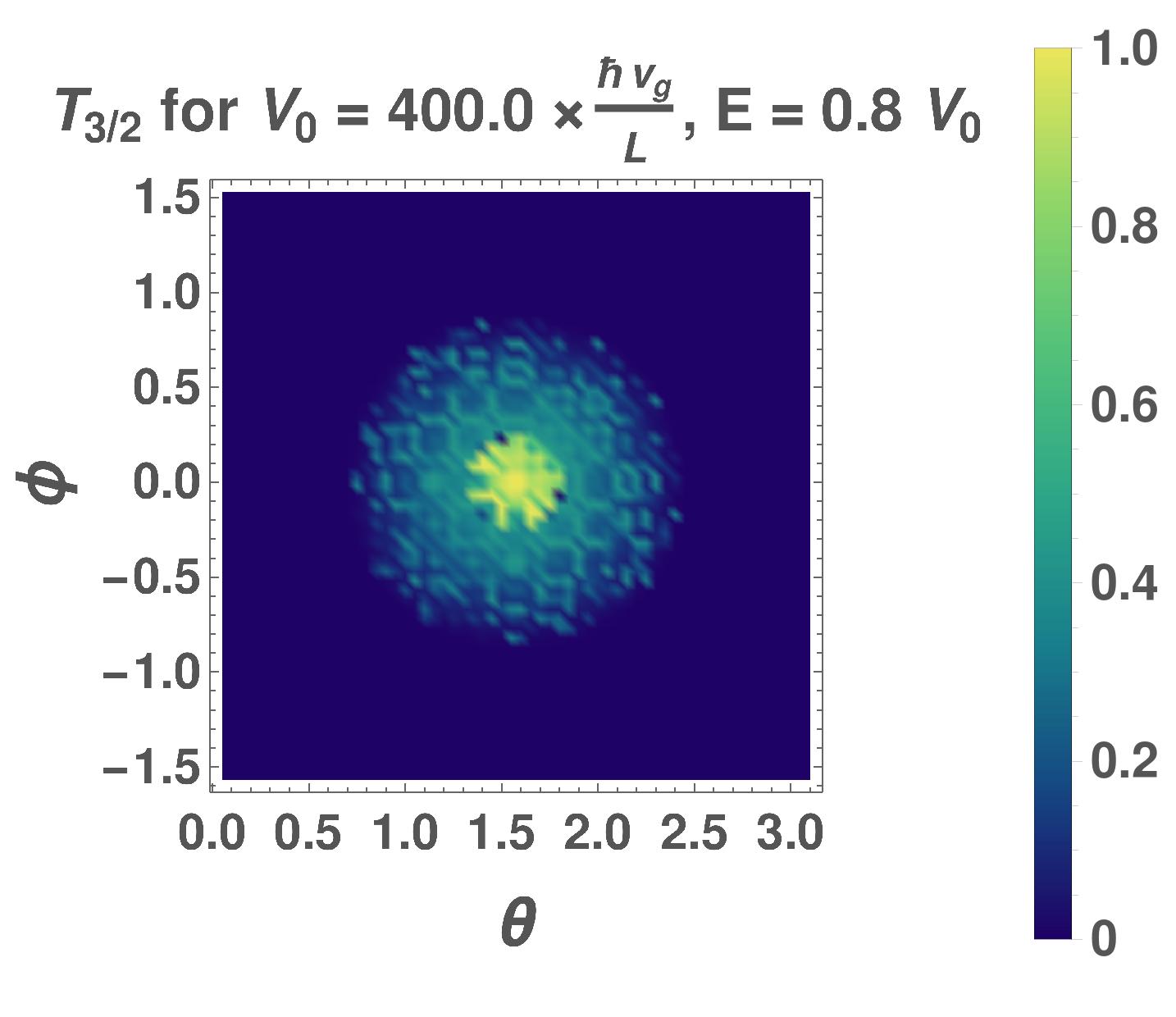}}\quad
\subfigure[]
{\includegraphics[width = 0.23 \textwidth]{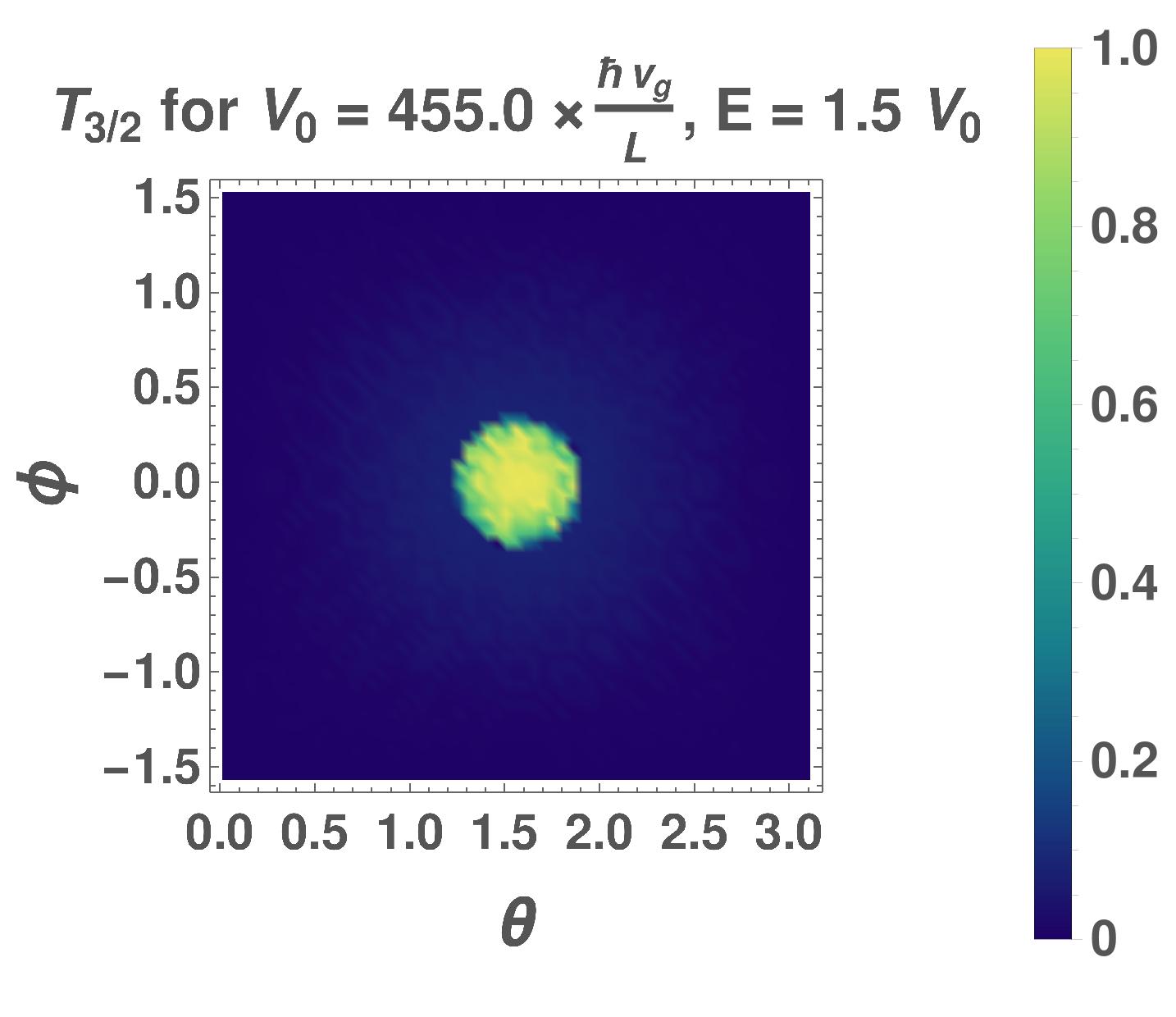}} 
\subfigure[]
{\includegraphics[width = 0.23 \textwidth]{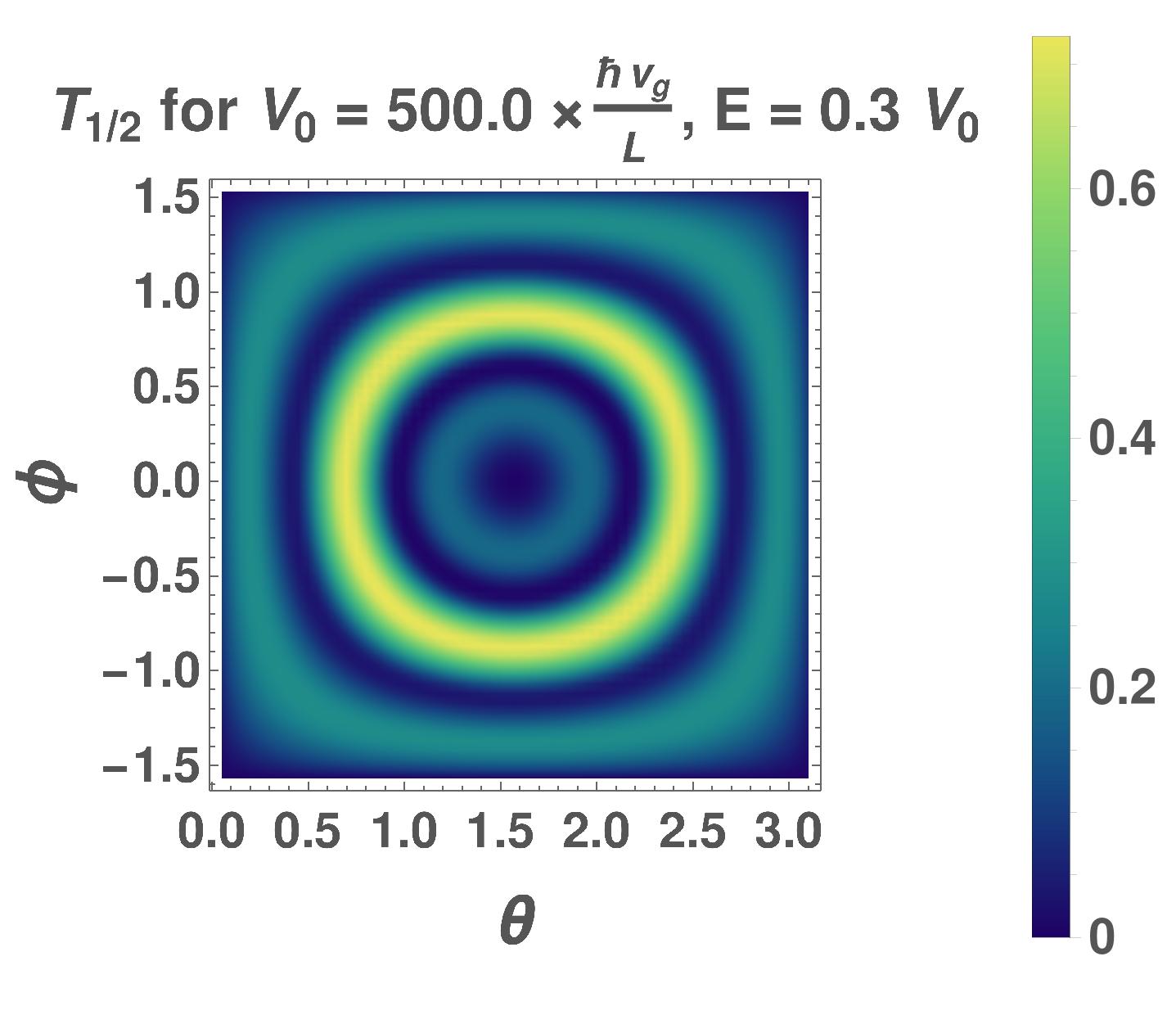}} \quad
\subfigure[]
{\includegraphics[width = 0.23 \textwidth]{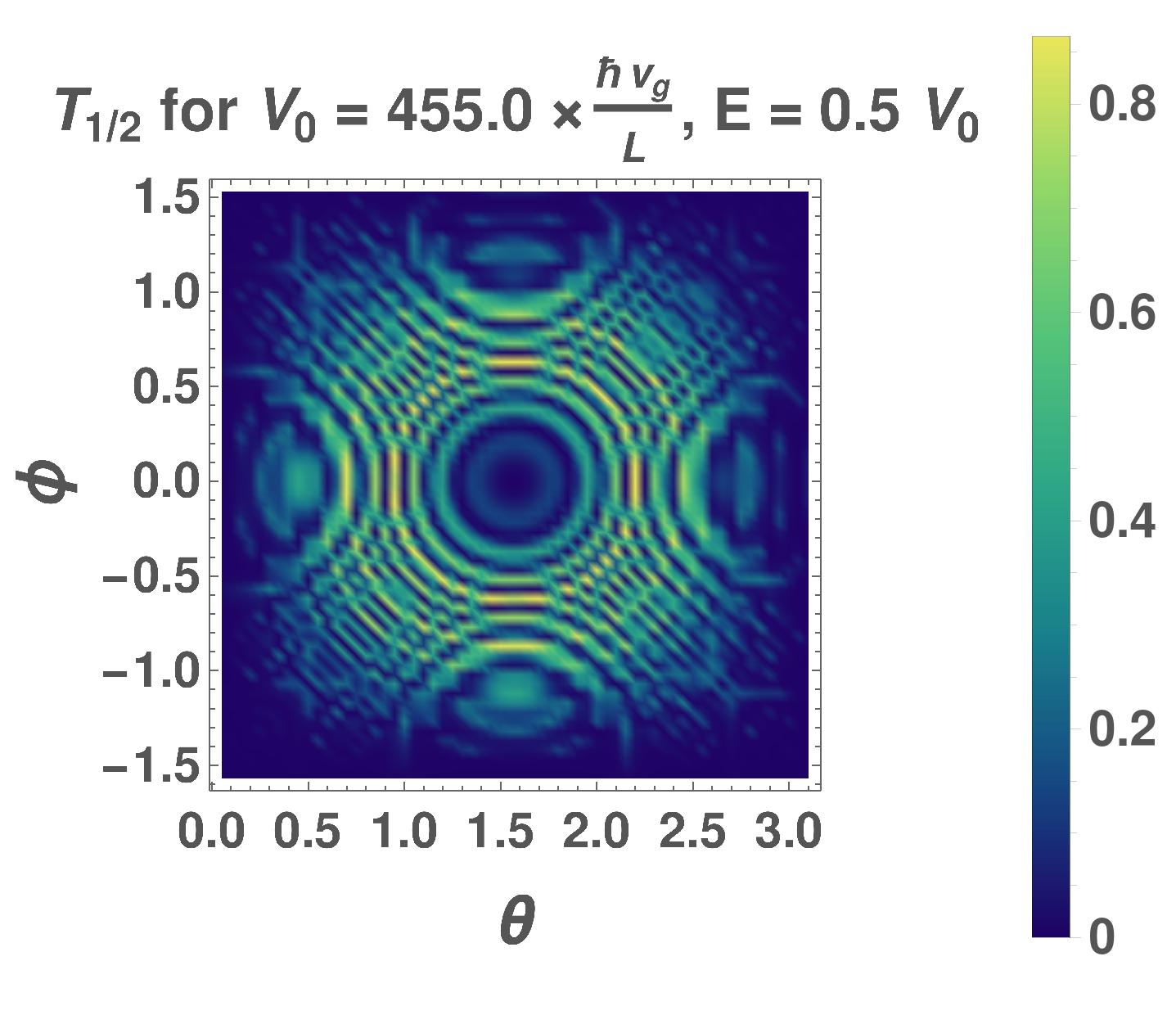}} \quad
\subfigure[]
{\includegraphics[width = 0.23 \textwidth]{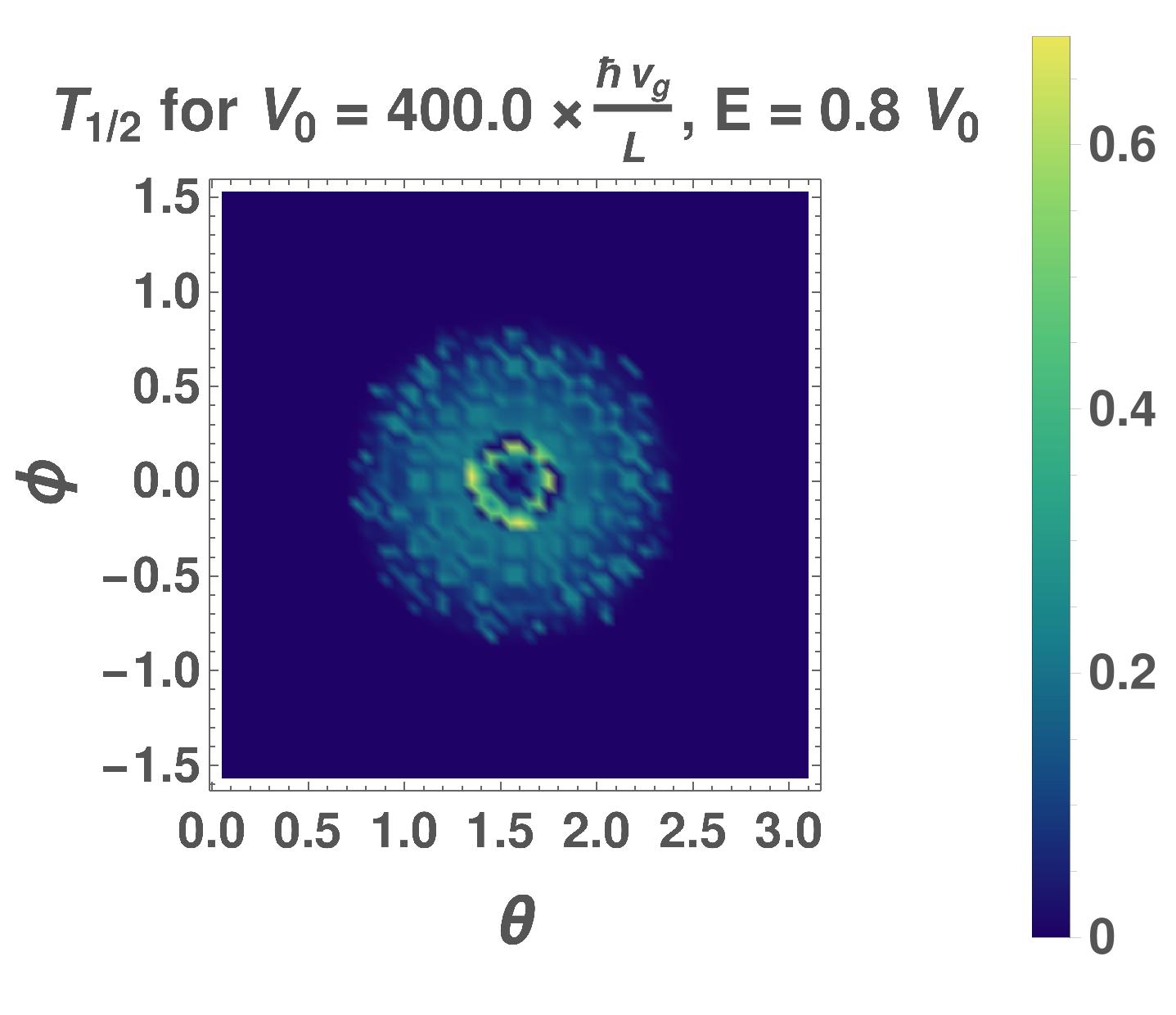}}\quad
\subfigure[]
{\includegraphics[width = 0.23 \textwidth]{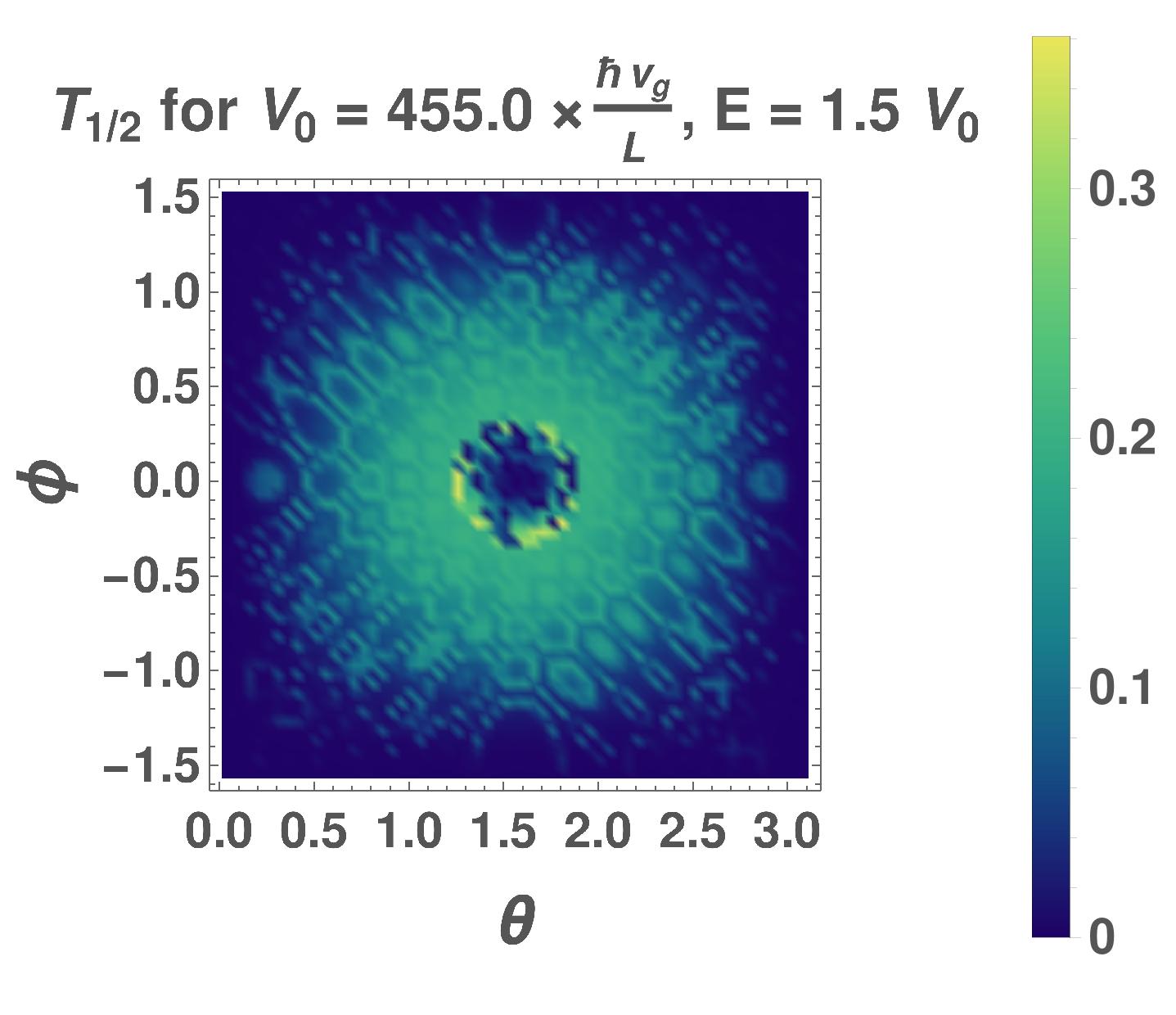}} 
\caption{Pseudospin-3/2 semimetal: Contourplots of the transmission coefficient ($T_\sigma $) in the absence of the vector potential, as functions of $(\theta, \phi)$, for various values of $V_0$ and $E$.
Klein tunneling is observed for a range of angles around normal incidence.}
\label{figcontourspin32}
\end{figure}
\subsection{Transmission coefficients, conductivity, and Fano factor}

\begin{figure}[h!]
\subfigure[]
{\includegraphics[width = 0.45 \textwidth]{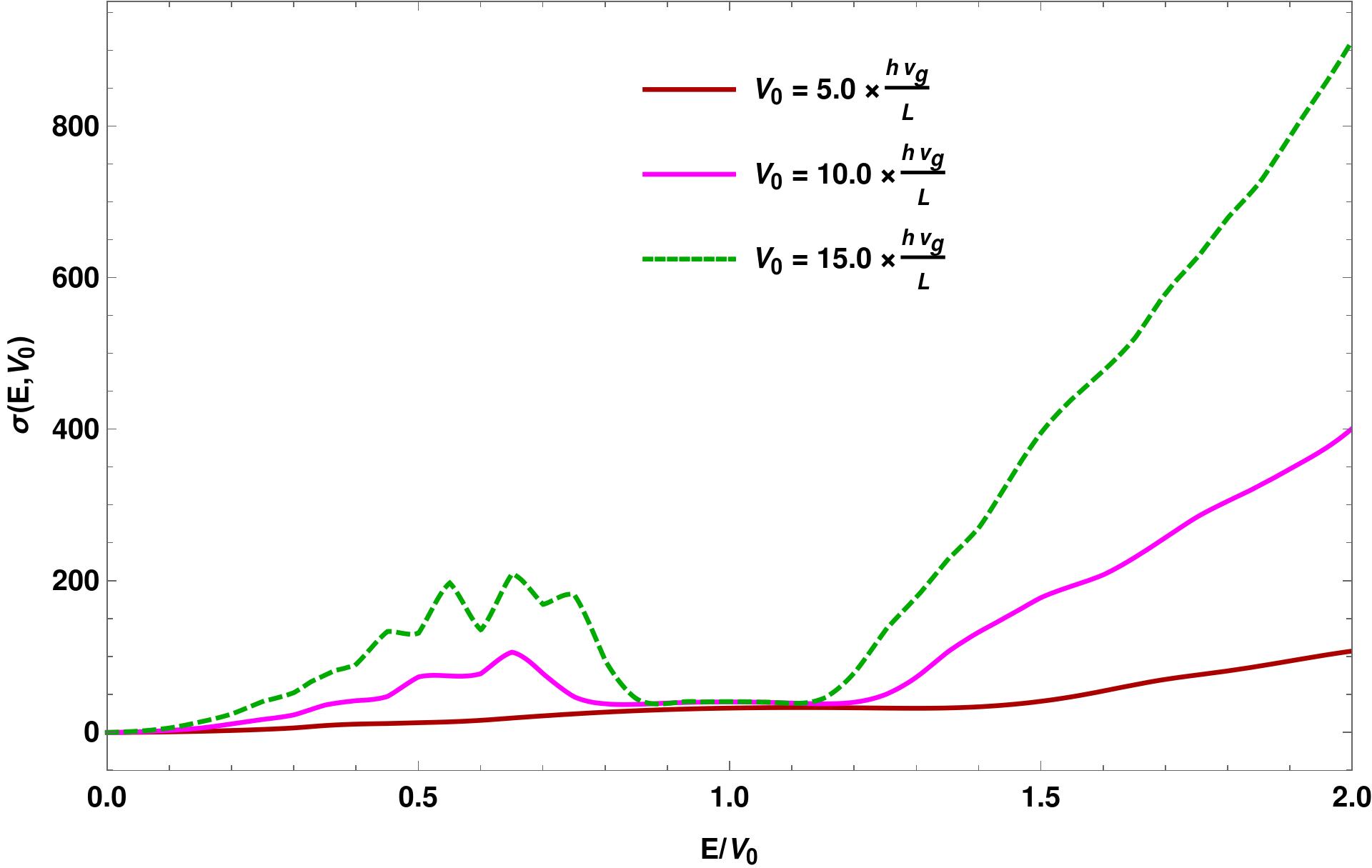}}\qquad
\subfigure[]
{\includegraphics[width = 0.45 \textwidth]{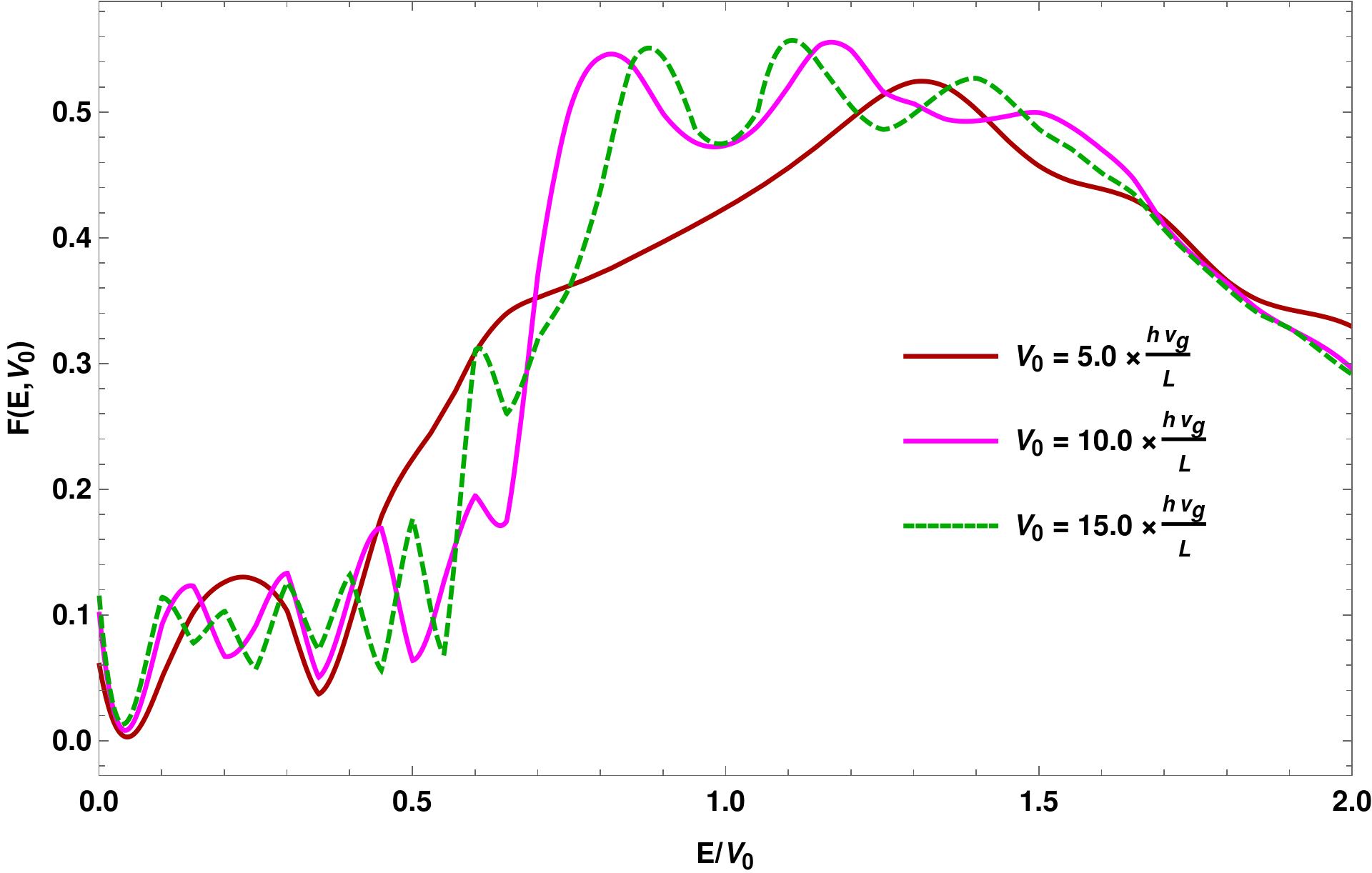}}
\caption{Pseudospin-3/2 semimetal: Plots of the (a) conductivity ($\sigma$ in units of $ 4/9$), and (b) Fano factor ($F$), as functions of $E/V_0$, for various values of $V_0$, in absence of the vector potential.}
\label{figfanospin32}
\end{figure}

\begin{figure}[htb]
\subfigure[]
{\includegraphics[width = 0.23 \textwidth]{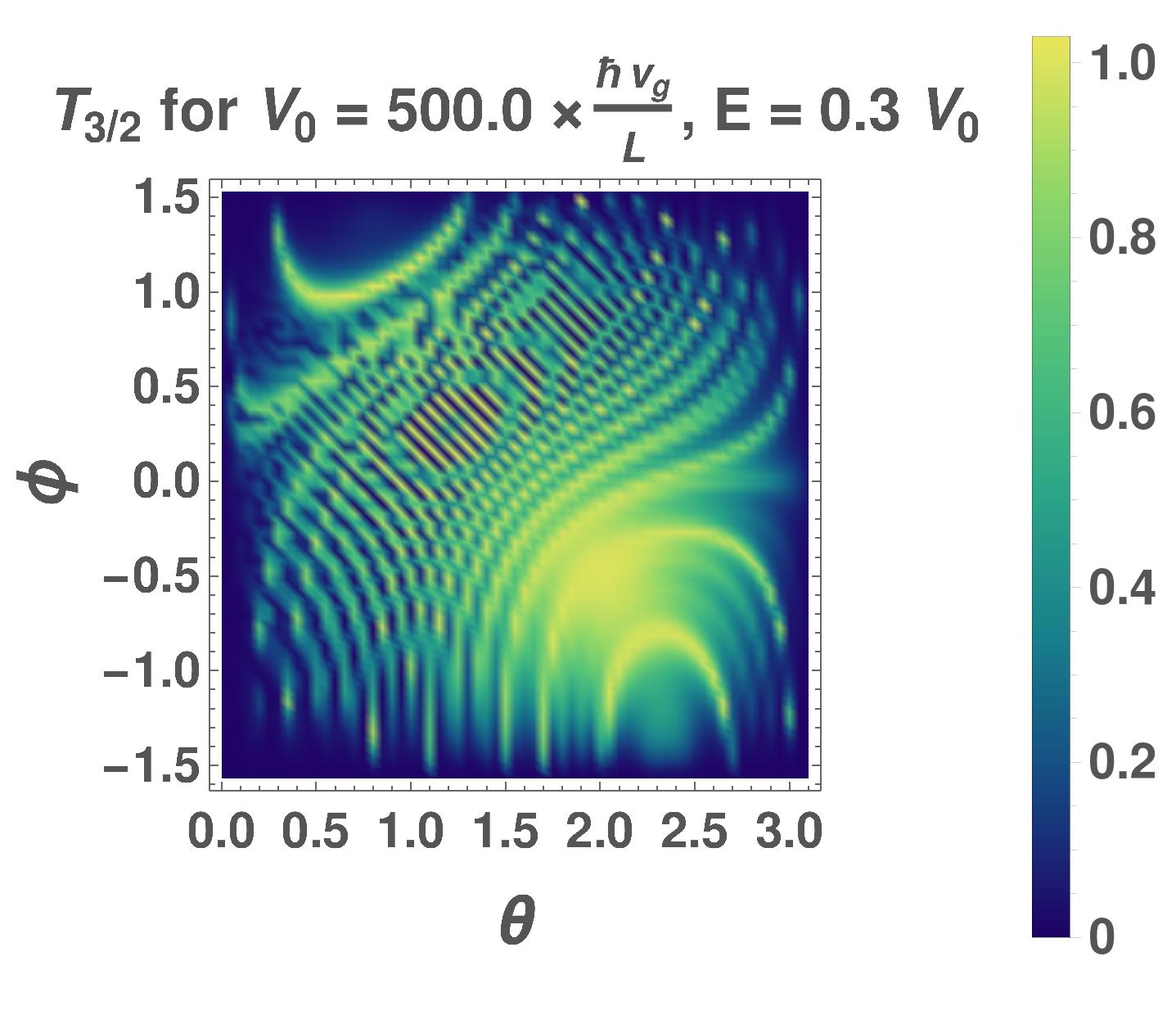}} \quad
\subfigure[]
{\includegraphics[width = 0.23 \textwidth]{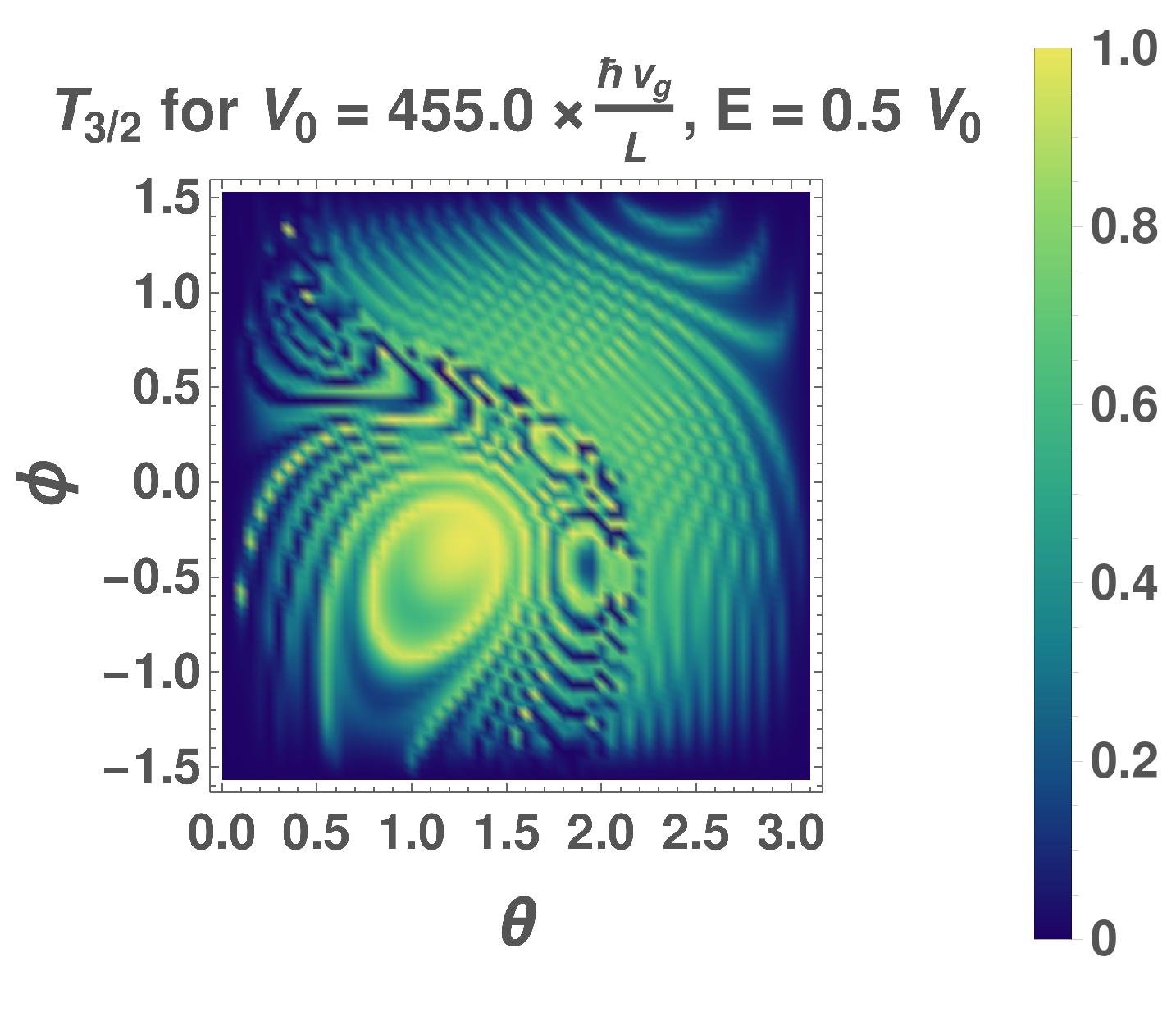}} \quad
\subfigure[]
{\includegraphics[width = 0.23 \textwidth]{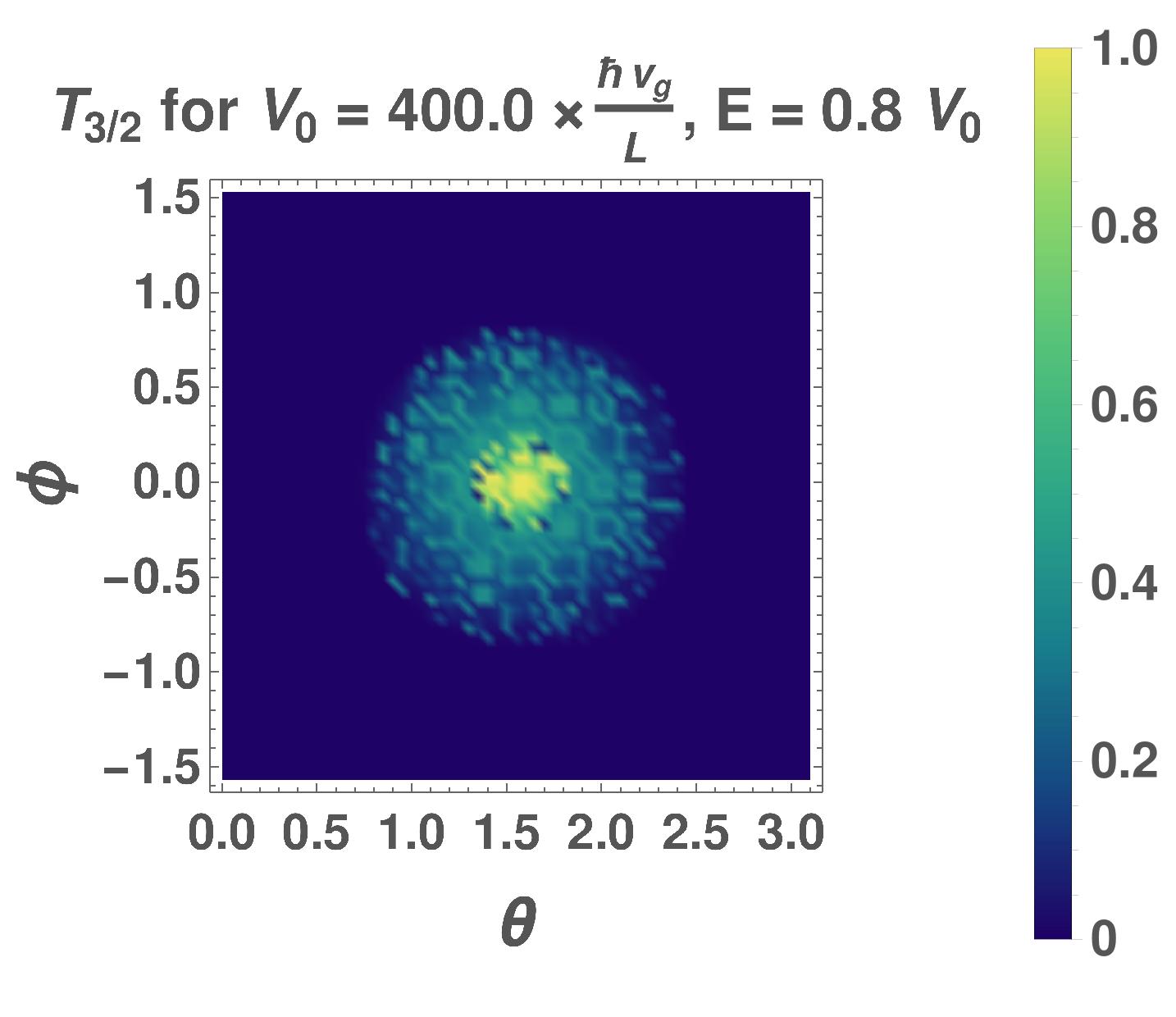}}
\subfigure[]
{\includegraphics[width = 0.23 \textwidth]{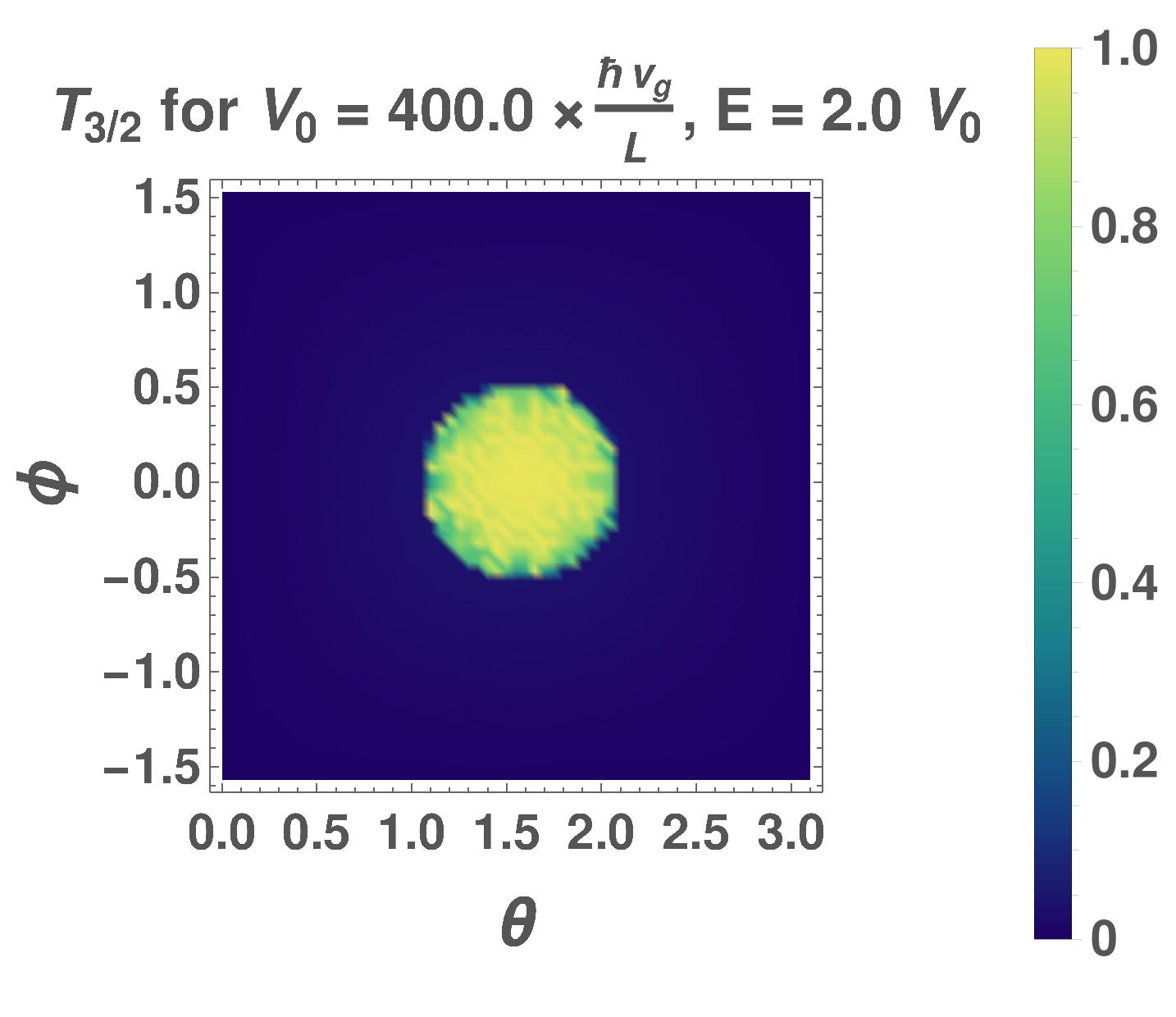}} 
\subfigure[]
{\includegraphics[width = 0.23 \textwidth]{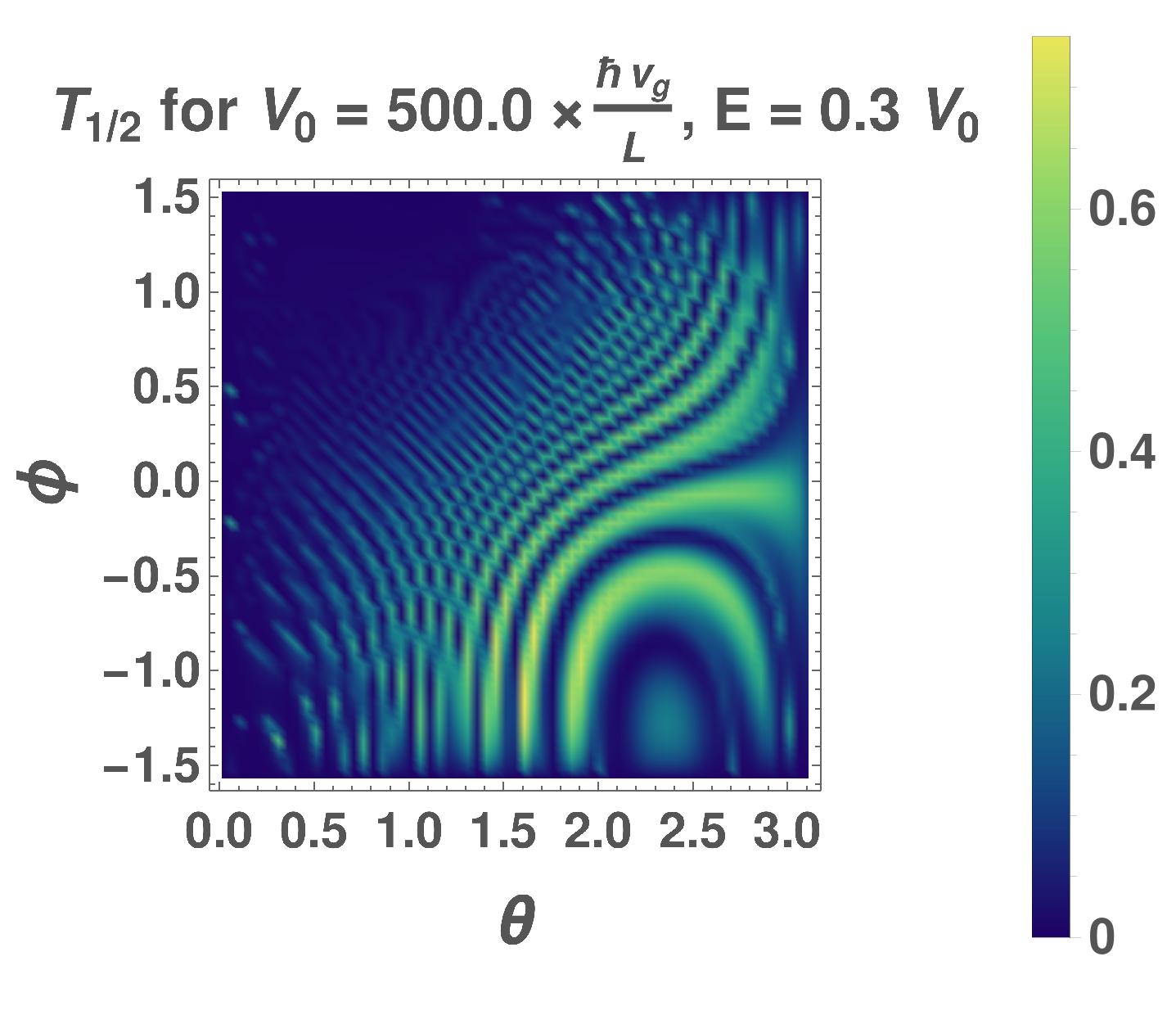}} \quad
\subfigure[]
{\includegraphics[width = 0.23 \textwidth]{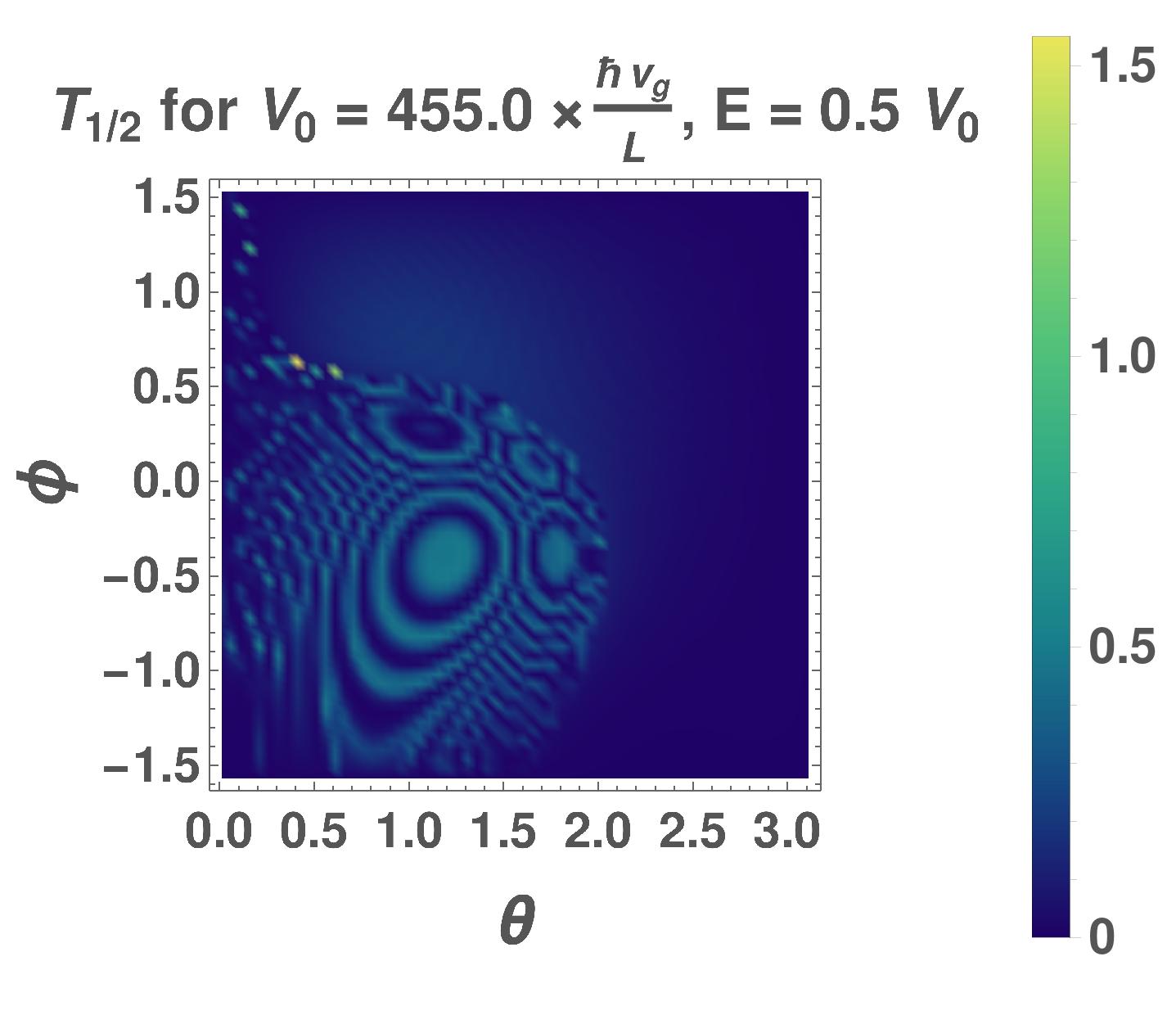}} \quad
\subfigure[]
{\includegraphics[width = 0.23 \textwidth]{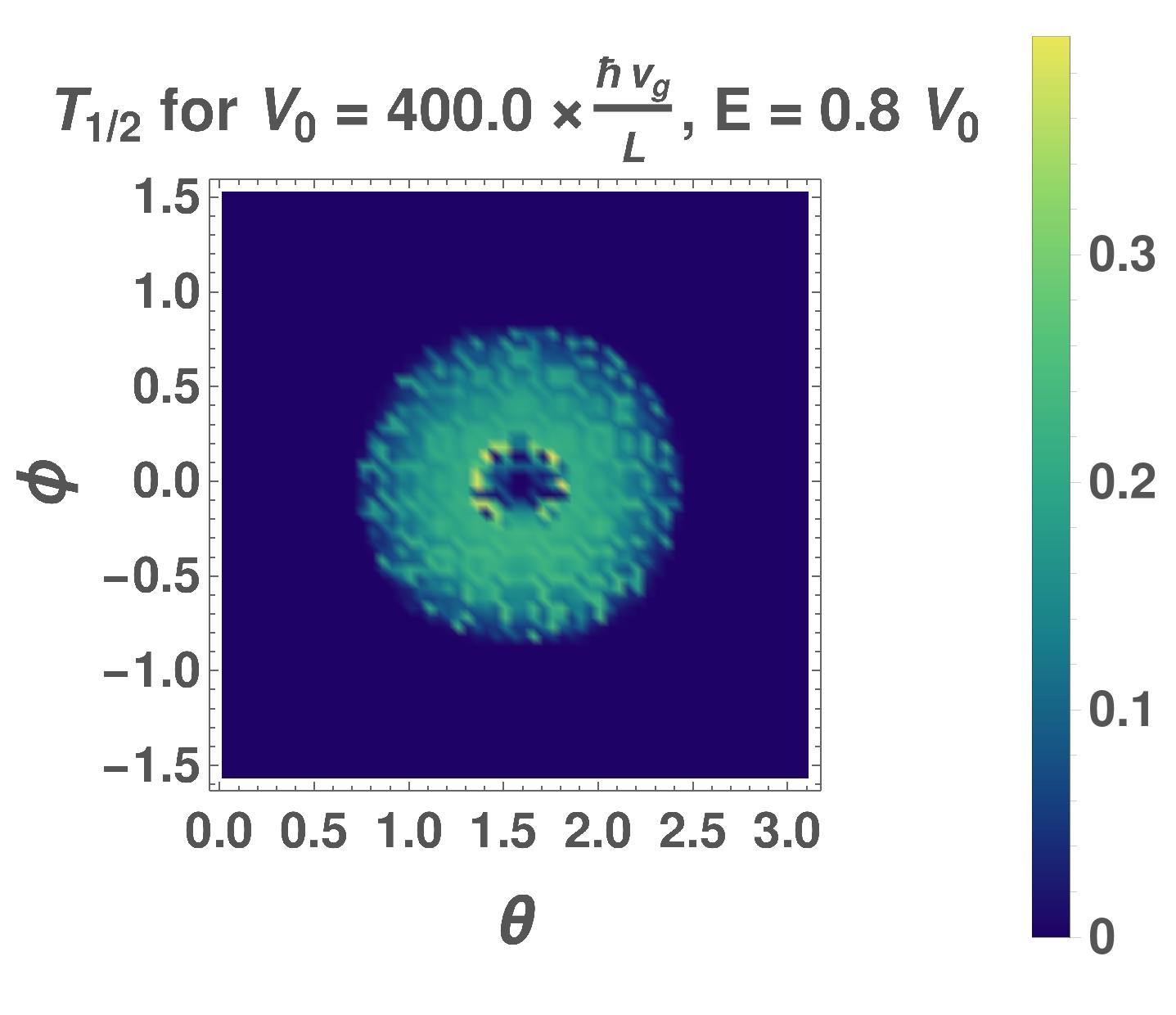}}\quad
\subfigure[]
{\includegraphics[width = 0.23 \textwidth]{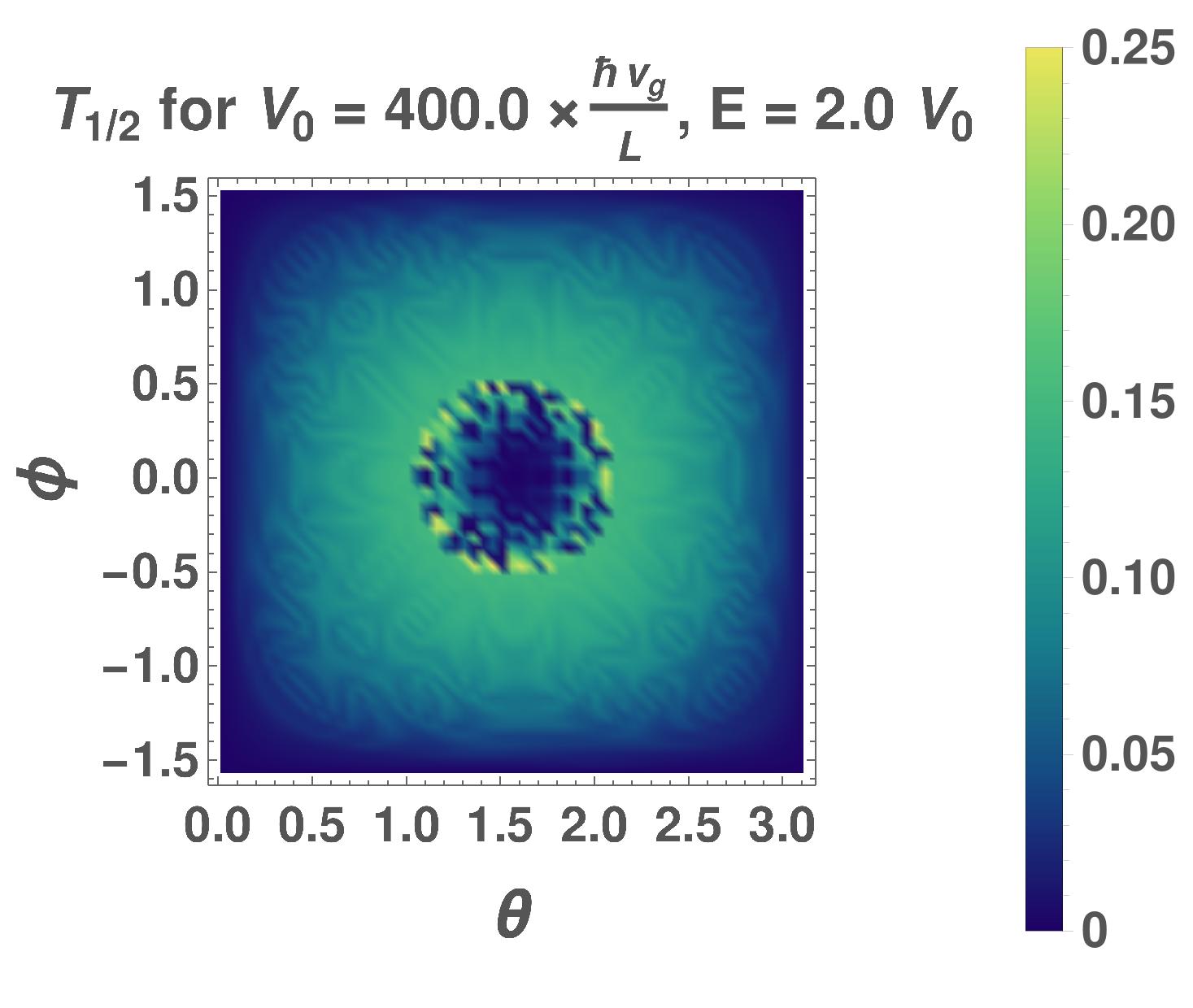}} 
\caption{Pseudospin-3/2 semimetal: Contourplots of the transmission coefficients ($T_\sigma$) in the presence of the vector potential, as functions of $(\theta, \phi)$, for various values of $V_0$ and $E$. The values for the vector potential components $\left \lbrace  A_y, A_z \right \rbrace $ are equal to: (a) and (e) $\left \lbrace 0.3,\,0.3\right  \rbrace  \frac{ V_0} {e\,v_g} $,
(b) and (f) $\left \lbrace 0.2,\,-0.2\right  \rbrace  \frac{ V_0} {e\,v_g}$,
(c) and (g) $\left \lbrace 0.01,\,0.01\right  \rbrace  \frac{ V_0} {e\,v_g}$,
(d) and (h) $\left \lbrace -0.01,\,0.01\right  \rbrace  \frac{ V_0} {e\,v_g}$.
}
\label{figcontourBspin32}
\end{figure}

In the continuum limit for the transverse momenta, using
$ k_{\ell,3/2} = \frac{2\,E} {3\,\hbar\,v_g} \sin \theta \cos \phi$,
$ n_y= \frac{2\,W E } { 3\, h\,v_g}   \sin \theta \sin \phi $,
$ n_z= \frac{ 2\,W E } { 3\,h\,v_g}   \cos \theta $, and
$ dn_y \, dn_z =
\frac{ 4\,W^2 E^2 } { 9\,h^2 v_g^2}  \cos \phi \,\sin^2 \theta \, d\phi \,d\theta $,
the conductance is given by \cite{blanter-buttiker}:
\begin{align}
G(E,V_0) & 
= \frac{4\, e^2\,W^2\, E^2 } { 9\,h^3 v_g^2} 
  \int_{ \theta=0 }^{\pi} \int_{\phi=-\frac{\pi}{2}}^{\frac{\pi}{2}}
\sum \limits_\sigma T_\sigma( E ,V_0,\theta, \phi,\mathbf B) \cos\phi \sin^2 \theta \,d\phi \,
d\theta,
\end{align}
leading to the conductivity expression:
\begin{align}
\sigma (E,V_0,\mathbf B)  & = \frac{4}{9} \left( \frac{ E } {h v_g/L} \right)^2
\int_{ \theta=0 }^{\pi} \int_{\phi=-\frac{\pi}{2}}^{\frac{\pi}{2}}
\sum \limits_\sigma T_\sigma( E ,V_0,\theta, \phi,\mathbf B) \cos\phi \sin^2 \theta \,d\phi\,
d\theta.
\end{align}
The Fano factor is given by:
\begin{align}
F(E,V_0,\mathbf B)  &=\frac  
{\int_{ \theta=0 }^{\pi} \int_{\phi=-\frac{\pi}{2}}^{\frac{\pi}{2}}
\sum \limits_\sigma T_\sigma( E ,V_0,\theta, \phi,\mathbf B) 
 \left[1-T_\sigma( E ,  V_0, \theta, \phi,\mathbf B) \right]
\cos\phi \sin^2 \theta \,d\phi \,d\theta  }
{ \int_{ \theta=0 }^{\pi} \int_{\phi=-\frac{\pi}{2}}^{\frac{\pi}{2}}
\sum \limits_\sigma T_\sigma( E ,V_0,\theta, \phi,\mathbf B) \cos\phi \sin^2 \theta \,d\phi
\,d\theta } \,.
\end{align}

In the absence of the magnetic fields,
Fig.~\ref{figpolarspin32} shows the polar plots of the two transmission coefficients as functions of the incident angle $\phi$, for the case when $k_z=0$ (hence $\theta=\pi/2$). Klein tunneling is observed for $T_{3/2}$ for a range of angles around normal incidence (independent of the incident energy).
Additionally, there are resonance conditions for certain values of $\tilde k$ and $L$ under which the barrier becomes completely transparent for $T_{3/2}$ or $T_{1/2}$. 
Fig.~\ref{figcontourspin32} shows the angular dependence of $T( E , V_0,\theta,\phi,\mathbf 0)$ in contourplots. The patterns for $E=V_0/2$ clearly distinguish the pseudospin-3/2 semimetal from the pseudospin-1 case, as the latter exhibits perfect transmission at all angles for $E=V_0/2$.
Klein tunneling is observed  for $T_{3/2}$ for a range of angles around $(\theta=0, \phi=0)$, and in those regions, $T_{1/2} = 0$ (since $ T_{3/2} + T_{1/2} \leq 1$).
In Fig.~\ref{figfanospin32}, we illustrate the conductivity $\sigma (E,V_0,\mathbf 0)$ and the Fano factor $ F (E,V_0,\mathbf 0)$, as functions of $E/V_0$, for some values of $V_0$. Unlike the pseudospin-1 case,
$F$ does not go to zero at $E=V_0/2$, due to the absence of super-Klein tunneling. Both $F$ and $\sigma$ show much more oscillatory behaviour compared to the pseudospin-1 quasiparticles.

The contourplots in Fig.~\ref{figcontourBspin32} capture how the presence of the vector potential modifies the transmission coefficients. The angles for perfect transmission is now shifted away from normal incidence.
The transmission patterns are also markedly different from those for the pseudospin-1 semimetals, as seen by comparing with Fig.~\ref{figcontourBspin1}.

\section{Summary and Discussions}
\label{secsum}

In this paper, we have computed the transmission coefficients of the pseudospin-1 and pseudospin-3/2 semimetals with linear dispersion and nonzero Chern numbers. These are the higher-pseudospin generalizations of the well-studied Weyl semimetals. The transmission coefficients have been calculated in the presence of both scalar and vector potentials, existing uniformly in a bounded region. The patterns found clearly serve as fingerprints of the corresponding semimetal, although all of them have linear dispersions. Similar computations were done for the case of Weyl fermions in Ref.~\cite{mansoor}. Comparing with those results, one can easily see that the characteristics for these higher-pseudospin cases differ considerably. In particular, the pseudospin-1 case demonstrates super-Klein tunneling, which is absent in the Weyl and pseudospin-3/2 cases.
The conductivities and Fano factors obtained here also serve as another set of measurable quantities to identify the different types of semimetals. Another important point is that this kind of calculations will help us find the perfect transmission regions by tuning the Fermi level and/or the magnetic fields, which has the potential to be used in generating localized transmission in the bulk of
the semimetals, for example in electro-optic applications.

The behaviour of the quantities calculated here can also be contrasted against that in quadratic band-crossing semimetals studied in Ref.~\cite{ips-qbcp-tunnel}.
In future works, these transport properties will be studied in the presence of disorder, as has been done in the case of Weyl \cite{emil2} and double-Weyl \cite{emil} nodes. The effect of magnetic fields
on the tunneling behaviour of the 3d double-Weyl nodes and 2d anisotropic Weyl fermions \cite{ips-kush} will be another interesting avenue to explore.
Furthermore, it will be worthwhile to examine the effects of terms which reduce the symmetry. For example, addition of a term proportional to $k_i\,J_i^3 $ to $ \mathcal{H}_{3/2}$ reduces the full rotational symmetry to the rotational cubic group. Lastly, this exercise needs to be carried out in the presence of interactions, which
can destroy the quantization of various physical quantities in the topological phases \cite{kozii,Mandal_2020}.

\section{Acknowledgments}
We thank Atri Bhattacharya for help with the figures.

\bibliography{biblio}
\end{document}